\documentclass{aa}  
\usepackage{CJK}
\usepackage{graphicx}
\usepackage{subfigure}
\usepackage{float}
\usepackage{amsmath}    
\usepackage{amssymb}     
\usepackage[normalem]{ulem}
\usepackage{lineno}
\usepackage{color}

\usepackage{txfonts}
\begin{document} 

\begin{CJK*}{UTF8}{gbsn}

   \title{Multiwavelength radiation from the interaction between magnetar bursts and a companion star in a binary system}
   
   \titlerunning{Radiation in Magnetar Binary System}
   \authorrunning{Wei et al.}

   \author{Yu-Jia Wei (魏煜佳)\inst{1, 2}
        \and
            Yuan-Pei Yang (杨元培)\inst{3,4}
            \and
        Da-Ming Wei (韦大明)\inst{1,2}
        \and
        Zi-Gao Dai (戴子高)\inst{5}
        }

    \institute{Key Laboratory of Dark Matter and Space Astronomy, Purple Mountain Observatory, Chinese Academy of Sciences, Nanjing 210023, China\\
        \email{dmwei@pmo.ac.cn}
        \and
            School of Astronomy and Space Science, University of Science and Technology of China, Hefei 230026, China
        \and
        South-Western Institute for Astronomy Research, Yunnan University, Kunming 650504, China\\
        \email{ypyang@ynu.edu.cn}
        \and
        Purple Mountain Observatory, Chinese Academy of Sciences, Nanjing 210023, China
        \and
        Department of Astronomy, School of Physical Sciences, University of Science and Technology of China, Hefei 230026, China;\\      \email{daizg@ustc.edu.cn}
    }

   \date{Received September 15, 1996; accepted March 16, 1997}

  \abstract
   {Magnetars are young, highly magnetized neutron stars that are associated with magnetar short bursts (MSBs), magnetar giant flares (MGFs), and at least some fast radio bursts (FRBs). In this work, we consider a magnetar and a main sequence star in a binary system and analyze the properties of the electromagnetic signals generated by the interaction between the magnetar bursts and the companion star. During the preburst period, persistent radiation could be generated by the interaction between the $e^+e^-$-pair wind from the magnetar and the companion or its stellar wind. We find that for a newborn magnetar, the persistent preburst radiation from the strong magnetar wind can be dominant, and it is mainly at the optical and ultraviolet (UV) bands. 
   For relatively old magnetars, the re-emission from a burst interacting with the companion is larger than the persistent preburst radiation and the luminosity of the companion itself. The transient re-emission produced by the heating process has a duration of $0.1 - 10^5 {\rm~s}$ at the optical, UV, and X-ray bands. 
   Additionally, we find that if these phenomena occur in nearby galaxies within a few hundred kiloparsecs, they could be detected by current or future optical telescopes.}


   \maketitle
%
\section{Introduction}

A magnetar is a type of young neutron star (NS) with an extremely strong magnetic field, which can reach strengths of $\sim 10^{15} {\rm~G}$~\citep[e.g.,][for a review]{Mereghetti_2008A&ARv..15..225M,Kaspi_2017ARA&A..55..261K}. The decay of the strong magnetic field can power strong emission, which is typically at X-ray and soft $\gamma$-ray bands with a duration from a few milliseconds to several months. 
The concept of the magnetar was first postulated for the pulsating gamma-ray burst (GRB) observed on March 9, 1979, in the Large Magellanic Cloud (LMC)~\citep[e.g.,][]{Mazets_1979Natur.282..587M,Daugherty_1983ApJ...273..761D,Usov_1984Ap&SS.107..191U,Dermer_1990ApJ...360..197D}. 
With more observations including a regular 8~s oscillating decay and its localization~\citep{Cline_1980BAAS...12..448C,Cline_1982ApJ...255L..45C}, this event was later considered as a soft gamma repeater (SGR) rather than a typical GRB.
Additionally, magnetars were then unified with anomalous X-ray pulsars (AXPs) by Thompson and Duncan~\citep{Thompson_1993ApJ...408..194T,Thompson_1996ApJ...473..322T}.
Observations have revealed many similarities between AXPs and SGRs~\citep{Mereghetti_2008A&ARv..15..225M}.
Currently, there are nearly 30 candidate magnetars discovered in the Milky Way, as summarized in~\cite{Kaspi_2017ARA&A..55..261K}. 

Evidence for the existence of younger magnetars has been found in recent years \citep[e.g.,][]{Kaspi_2017ARA&A..55..261K}.
A very young radio-loud magnetar, Swift J1818.0-1607, was discovered in 2020
\citep{Esposito_2020ApJ...896L..30E}. It has a spin period of $1.36 \rm~s$ and a characteristic age of 240 yr, and is presently the youngest known magnetar in the Milky Way. 
\cite{Hu_2020ApJ...902....1H} further revised the age of Swift J1818.0-1607 to $\sim 470$~yr based on the observation by the Neutron star Interior Composition Explorer (NICER) team.
Some high-energy outbursts from extragalactic magnetars have recently been observed. 
Several GRB-like events have been proposed to be actually magnetar giant flares (MGFs) after estimating their distances. MGFs are characterized by a typical duration of $\sim 0.1 {\rm~s}$ and an isotropic energy of $\sim 10^{45}-10^{46} {~\rm erg}$ at the hard X-ray and soft $\gamma$-ray bands~\citep[e.g.,][]{Burns_2021ApJ...907L..28B}. For example, GRB 200415A is considered to be an extragalactic MGF with an isotropic energy of $\simeq 1.36 \times 10^{46} ~{\rm erg}$ and a distance of $\simeq 3.5 {\rm~Mpc}$~\citep[e.g.,][]{Minaev_2020AstL...46..573M,Yang_2020ApJ...899..106Y,Roberts_2021Natur.589..207R,Svinkin_2021Natur.589..211S}. \cite{Zhang_2022arXiv220507670Z} proposed that the age of this magnetar is constrained to only a few weeks by interpreting the period in the decaying light curves as the rotation period.
Recently, GRB 231115A, detected by INTEGRAL in M 82,  has also been considered to be another possible extragalactic MGF~\citep{Wang_2023arXiv231202848W,Mereghetti_2023arXiv231214645M}.
In addition, extremely young magnetars have been suggested to be the central engines of GRBs and supernovae.
The prolonged plateaus in the X-ray afterglows of some GRBs have been suggested to be mainly contributed by the energy injection from extremely young magnetars~\citep[e.g.,][]{Kouveliotou_1998Natur.393..235K,Stratta_2018ApJ...869..155S}. A similar mechanism involving extremely young magnetars can also explain the light curves of supernovae, in which the radiation from the radioactive decay of Ni is not enough to provide the observed luminosity~\citep[e.g.,][]{Kasen_2010ApJ...717..245K}. 

Magnetars have also been confirmed as the central engines of at least some of the recorded fast radio bursts (FRBs). In 2020, a Galactic FRB 200428 with an isotropic energy of $\sim 10^{35} ~\rm erg$~\citep{Pavlovi_2013ApJS..204....4P,Zhong_2020ApJ...898L...5Z,Zhou_2020ApJ...905...99Z,Yang_2021ApJ...919...89Y} detected by CHIME~\citep{CHIME_2020Nature..587..54} and STARE2~\citep{Bochenek_2020} was confirmed as being related to the magnetar SGR 1935+2154, which has a characteristic age of about $3.6 {\rm~kyr}$~\citep{Israel_2016MNRAS.457.3448I}. This indicates that magnetars are at least one of the physical origins of FRBs. In addition, FRB 200428 was found to be associated with a magnetar short burst (MSB) of $\sim 0.5 {\rm~s}$ in duration and  $\sim 10^{40} {\rm~erg}$~in isotropic energy \citep{Tavani_2021NatAs...5..401T}, which shows magnetars can produce both MSBs and FRB-like radio bursts simultaneously. SGR 1935+2154 is regarded as one of the most burst-active SGRs~\citep{Hess_2022icrc.confE.777H}, emitting 127 bursts from 2014 to 2016~\citep{Lin_2020ApJ...893..156L}. In late April and May 2020, SGR 1935+2154 re-entered an active phase, and more than 217 bursts were detected~\citep{Younes_2020ApJ...904L..21Y}. Many more bursts have been observed from this source since then, including a large episode in October, 2022~\citep{Younes_2022ATel15674....1Y,Maan_2022ATel15697....1M, Palmer_2022ATel15752....1P}.

Additionally, young magnetars with a strong magnetic field and a fast rotation spin can produce an ultrarelativistic $e^+e^-$-pair wind~\citep[e.g.,][]{Rees_1974MNRAS.167....1R,Kennel_1984ApJ...283..694K,Coroniti_1990ApJ...349..538C,Kirk_2003ApJ...591..366K}, which could produce various observable features.
Some studies have predicted radiation associated with such a pair wind. For example, \cite{Dai_2004ApJ...606.1000D} and \cite{Geng_2016ApJ...825..107G} studied the afterglow emission generated by the interaction of such a wind with a fireball from the GRB. If the magnetar is in a binary system, the radiation can be generated by the bow shock from the interaction between a magnetar $e^+e^-$-pair wind and a stellar wind~\citep[e.g.,][]{Cant_1996ApJ...469..729C,Wilkin_1996ApJ...459L..31W,Bucciantini_2005A&A...434..189B,Wadiasingh_2017ApJ...839...80W}. 
In addition to the ultrarelativistic pair wind, \citet{Yang_2021ApJ...920...34Y} proposed that FRBs from a magnetar can heat the companion in a binary system and produce the re-emission at the optical band with a duration of a few hundred seconds.
Some previous works have also studied the interaction process between the ejecta and the companion if a catastrophic event occurs in a binary system. The companion star can also have a shielding effect on the radiation from the burst~\citep{Zou_2021ApJ...921....2Z}, and \cite{MacFadyen_2005astro.ph.10192M} studied the X-ray flares generated by the interaction between the GRB outflow and a stellar companion with a relatively small distance. Furthermore, the radiation from the interaction of supernova ejecta and a companion star has also been studied~\citep[e.g.,][]{Kasen_2010ApJ...708.1025K}.

It is common for stars to be in binary or even multiple star systems~\citep[e.g.,][]{Duch_2013ARA&A..51..269D,Badenes_2018ApJ...854..147B}. 
There have been several studies addressing the possibility of magnetars existing in binary systems~\citep{Chrimes_2022MNRAS.513.3550C, Bozzo_2022MNRAS.510.4645B, Xu_2022RAA....22a5005X}. 
According to the simulation of \cite{Xu_2022RAA....22a5005X}, there are probably magnetars in high-mass X-ray binaries. Observationally, \cite{Bozzo_2022MNRAS.510.4645B} proposed that the X-ray binary 3A 1954+319 is likely a system with a magnetar and an M supergiant. Also, \cite{Beniamini_2023MNRAS.520.1872B} studied a list of magnetar candidates in binaries.
For the binary evolution process, several authors have studied the effects of supernova ejecta on binary star systems, and these studies found that the binary system can remain bound~\citep{Hirai_2018ApJ...864..119H}.
Observationally, \cite{Chen_2024Natur.625..253C} reported a 12.4 day periodicity in SN 2022jli, which indicates the orbital period of a close binary system after a supernova.
In addition to the classical binary evolution process, the formation of binary systems also includes the dynamic capture process. \cite{Lee_2010ApJ...720..953L} proposed that the binary system including compact objects can also be generated by the dynamic capture in dense stellar environments. In this way, the prior activity from the magnetar, including the supernova ejecta, will not destroy the binary system.

In this theoretical work, we consider a magnetar in a close binary system with a main sequence star as the companion. 
Once a burst (e.g., FRBs, MSBs, and MGFs) occurs on a magnetar, it will interact with a companion star due to its high radiation pressure. Here we mainly analyze three types of bursts: FRBs, MSBs, and MGFs. The blackbody radiation can be generated via bursts heating the companion star due to the optically thick companion surface. The ultrarelativistic $e^+e^-$-pair wind from the magnetar interacts with the companion star or its stellar wind and generates sustained radiation that we refer to as the persistent preburst radiation.
\cite{Bhardwaj_2021ApJ...910L..18B} reported a repeating FRB 20200120E associated with a globular cluster with an age of $\sim 9.13 {\rm~Gyr}$ in M 81, which is a spiral galaxy at $3.63 \pm 0.34 {\rm~Mpc}$, where the number of main sequence and NSs could be as large as $\gtrsim 10^2$~\citep{Kremer_2021ApJ...917L..11K}. Therefore, it is natural to consider the possibility of electromagnetic radiation being produced by a binary system with a magnetar, as analyzed by \cite{Yang_2021ApJ...920...34Y}  for the case of FRBs, and it is crucial that we improve our understanding of the observed features of FRBs and/or other various bursts (e.g., MSB and MGF) and their progenitors, considering the scarcity of confirmed observations of electromagnetic signals associated with FRBs. 

We calculate the persistent preburst radiation produced by the magnetar wind, and the re-emission from bursts interacting with the companion in Sect.~\ref{sect:judge} and Sect.~\ref{sect:burst}. 
We then calculate the spectra and the light curves for such radiation and consider its detectability with optical (the Vera C. Rubin Observatory) and X-ray telescopes (\textit{Einstein Probe}, \textit{Chandra} and \textit{XMM-Newton}) in Sect.~\ref{sect:lcandsp}.
Finally, in Sect.~\ref{sect:param}, we scan the parameter space to determine whether the radiation from the magnetar wind and/or the magnetar bursts might be obscured by the radiation from the companion star, and find the parameter space in which the re-emission from bursts dominates.

We use the notation $Q_x=Q/10^x$ in CGS units, except $t_{\rm yr}\equiv t/1~{\rm yr}$, $M_{\rm c,1 M_\odot}=M_{\rm c}/1 M_\odot$, $R_{\rm c,1 R_\odot}=R_{\rm c}/1 R_\odot,$ and $L_{\rm c,1 L_\odot}=L_{\rm c}/1 L_\odot$.

\section{Model}
\label{sect:model}

\begin{figure}
        \centering
    \includegraphics[width=0.4\textwidth]{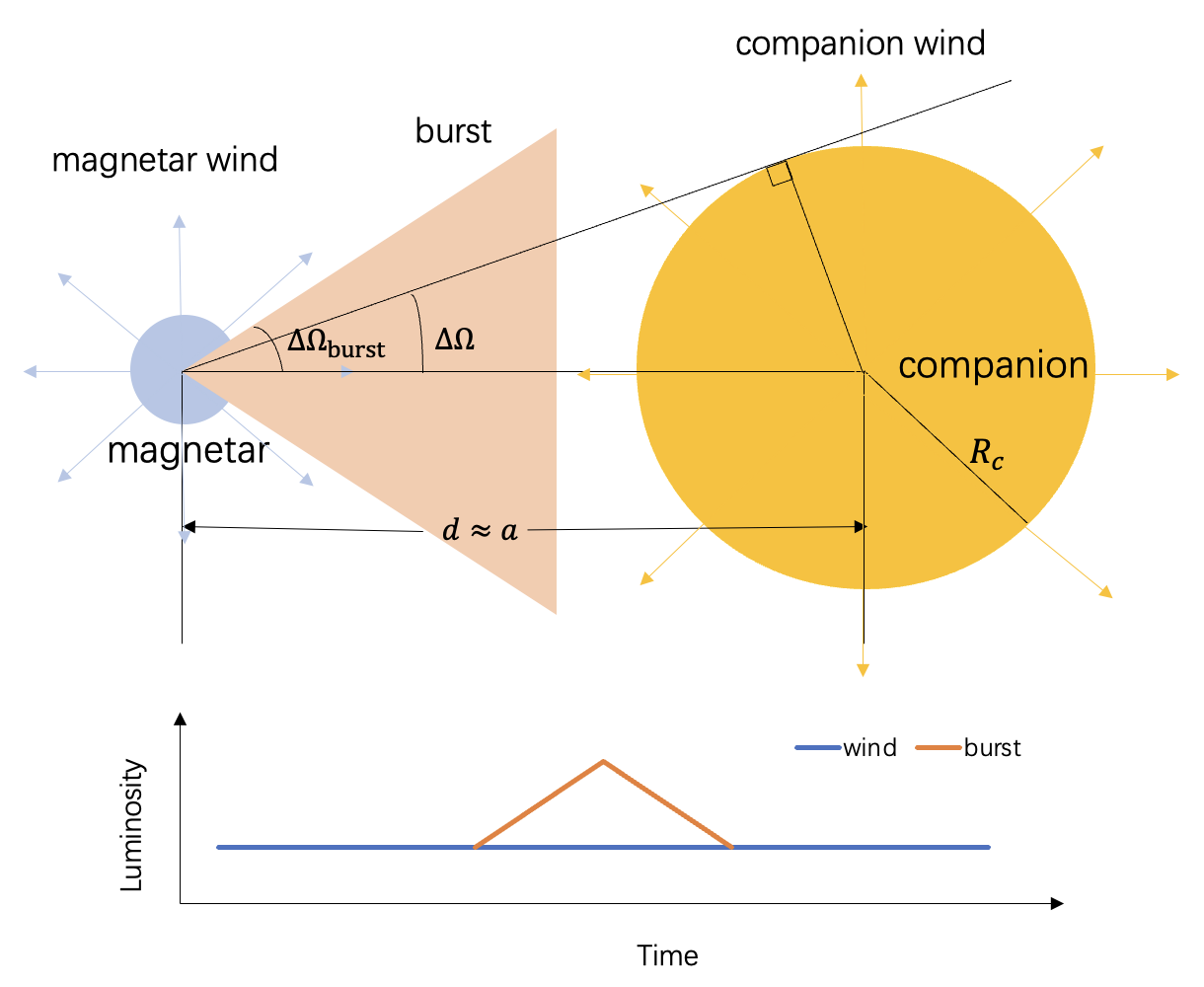}
        
        \caption{Schematic picture of radiation generated by a burst and/or a wind from a magnetar interacting with a main sequence star in a binary system. The persistent preburst radiation is generated by a strong magnetar wind heating a companion star, or by the bow shock from a weak magnetar wind interacting with a stellar wind from a companion star. The radiation from the transient is generated by the re-emission process from the companion star heated by a magnetar burst (e.g., an FRB, an MSB, and/or an MGF). }
        \label{fig:model}
\end{figure}

In this section, we discuss the physical processes of (1) the persistent preburst radiation from the interaction between the ultrarelativistic $e^+e^-$-pair wind from the magnetar and the companion or its stellar wind; and (2) the re-emission from the interaction between the magnetar bursts and the companion,  which appears to be transient relative to the persistent preburst radiation, as shown in Fig.~\ref{fig:model}.
In this case, the persistent preburst radiation is considered as the background before and after the generation of the re-emission from bursts, as shown in the bottom panel of Fig.~\ref{fig:model}. 
A magnetar is in a binary system with a main sequence star with $M_c \sim 0.1 - 20 {\rm~M_{\odot}}$. 
This radiation could be observable as long as the line of sight is not obstructed by the companion. 

\subsection{Persistent preburst radiation}
\label{sect:judge}

\begin{figure}
        \centering
        \subfigure[]{
                \begin{minipage}[b]{0.4\textwidth}
                        \begin{center}
                                \includegraphics[width=1.0\textwidth]{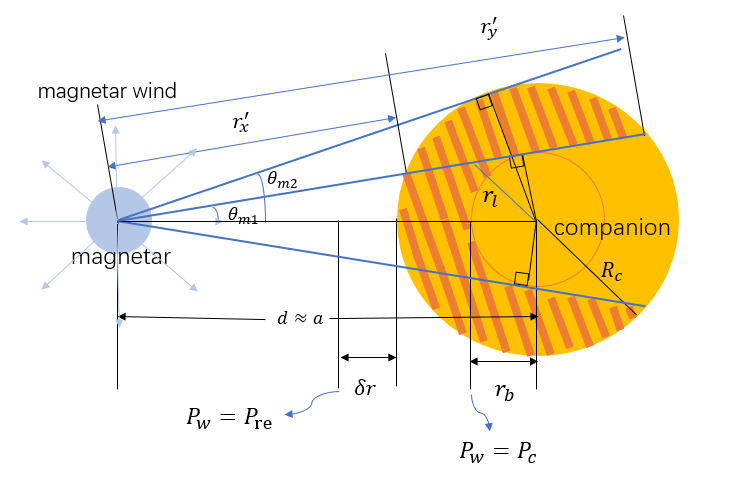}
                        \end{center}
                \end{minipage}   
                \label{fig:windmodel1}
        }%

        \subfigure[]{
                \begin{minipage}[b]{0.4\textwidth}
                        \begin{center}
                                \includegraphics[width=1.0\textwidth]{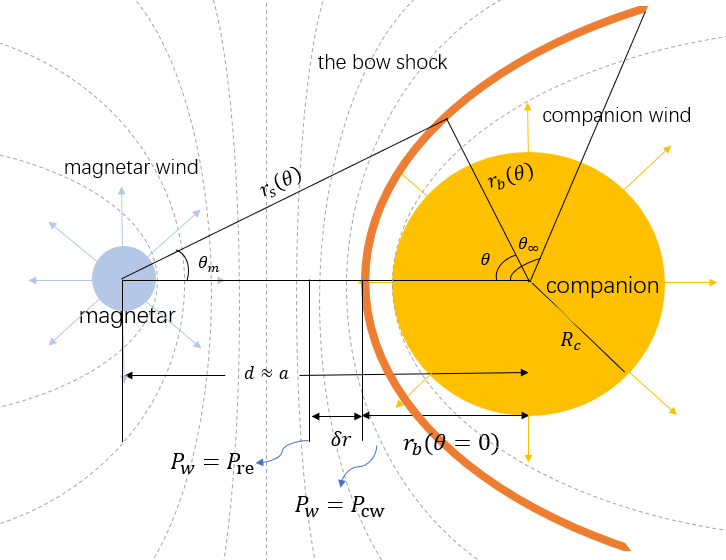}
                        \end{center}
                \end{minipage}
                \label{fig:windmodel2}
        }%
        \caption{Schematic picture of the persistent preburst radiation. In the top panel, we show the scenario (Case I) where the magnetar wind directly interacts with the companion star. In the bottom panel, we indicate the scenario (Case II) where the bow shock is generated by the interaction between the magnetar wind and the possible strong companion wind.}
        \label{fig:windmodel}
\end{figure}

The persistent preburst radiation can be divided into two cases based on the luminosity of the magnetar wind $L_w$, as shown in Fig.~\ref{fig:windmodel}. We refer to the scenario where the magnetar wind directly interacts with the companion star as Case I, while we refer to the scenario in which the magnetar wind interacts with the companion wind to produce a bow shock as Case II.

For a strong magnetar wind, it will directly interact with the companion star, as shown in Fig.~\ref{fig:windmodel1}. In this process, the bow shock will not be generated, because the pressure of the magnetar wind will be so high that the pressure equilibrium position of the two winds (i.e., the magnetar wind and the companion wind) will be inside the companion star. The magnetar wind will penetrate the companion and stop at $r_b$, where the pressure of the magnetar wind $P_w$ is equal to the internal pressure $P_c$ of the companion star. The energy from the magnetar wind will then be transferred to the swept medium of the companion star. We approximate that the pressure of the magnetar wind inside the companion star is a constant, because $d \gg R_c$, where $d$ is the distance between the magnetar and the companion star, and $R_c$ is the radius of the companion star. Therefore, the heated region of the companion is the shaded orange area in Fig.~\ref{fig:windmodel1}, and the re-emission can then be generated considering the black-body radiation. 

For a relatively weak magnetar wind, the bow shock will be generated by the interaction between the magnetar wind and the companion wind at $r_b$, where the momentum conservation is satisfied.
In Fig.~\ref{fig:windmodel2}, the dashed gray line shows the geometry of the bow shock with different $L_w$. The main radiation is contributed by the shock head considering the synchrotron process, and we assume the radiation region is a cone surface with a width of $\Delta_{\rm bow}$. 

\subsubsection{Criteria for distinguishing Case I from Case II}

In this section, we provide a detailed explanation of how to differentiate between Case I and Case II. For an extremely young magnetar with a millisecond spin period, the luminosity of a highly relativistic wind dominated by the energy flux of $e^+ e^-$ pairs is 
\begin{equation}
        L_w = L_{w, 0} \left(1 + \frac{T_{\rm age}}{T_{\rm sd} }\right)^{-2},
        \label{eq:Lw}
\end{equation}
where $L_{w,0} \simeq 9.6 \times 10^{44} {~\rm erg ~s^{-1}} B_{\perp, 15}^2 R_{M, 6}^6 P_{0,-2}^{-4}$ is the initial spin-down luminosity, $T_{\rm sd} \simeq 2.3 \times 10^5 {~\rm s} \ B_{\perp, 15}^{-2} I_{45} R_{M,6}^{-6} P_{0, -2}^2$ is the initial spin-down timescale of the magnetar, $B_\perp$ is the surface dipolar magnetic field of the magnetar, $I = 2 / 5 \left(M_m  R_m^2\right)$ is the moment of inertia of the magnetar, $R_m$ is the radius of the magnetar, $M_m$ is the mass of the magnetar, and $T_{\rm age}$ is the age of the magnetar. $P_0$ is the initial rotation period of the magnetar. The theoretical value of $P_0$ can be as low as $\sim 0.3 - 0.5$~ms~\citep{Cook_1994ApJ...422..227C, Koranda_1997ApJ...488..799K}. We note that the observed rotation period of the Galactic magnetar ranges from $0.32 - 12$~s~\citep{Archibald_2015ApJ...810...67A,Kaspi_2017ARA&A..55..261K,Kuiper_2018MNRAS.475.1238K}, and this is because of the magnetar spindown, which dictates that most Galactic magnetars with ages of a few hundred years to tens of thousands of years will have relatively long periods.
To be consistent with the observed persistent X-ray luminosity for Galactic magnetars~\citep{Olausen_2014ApJS..212....6O}, here we use $P_0 = 10^{-2} {\rm~s}$, $B_{\perp} = 10^{15} {\rm~G}$, $M_m = 1.4 {\rm~M_{\odot}}$, and $R_m = 10^6 {\rm~cm}$ unless stating otherwise.

The pressure of the relativistic wind should be
\begin{equation}
        \begin{aligned}
                P_{w} \simeq \frac{1}{3} U_{w} = \frac{L_w}{12 \pi (d - r)^2 c},
        \end{aligned}
        \label{eq:Pw}
\end{equation}
where $r$ is the distance between the area we study and the companion, as shown in Fig.~\ref{fig:windmodel}. We use the semimajor axis $a$ of the binary system to approximate $d$ (which applies to binary systems with a relatively small eccentricity), that is 
\begin{equation}
        a = {\left( \frac{G M_{\rm tot} P_{\rm orb}^2}{4 \pi^2} \right)}^{1/3},
\end{equation}
where $M_{\rm tot} = M_c + M_m$ is the total mass of the binary system, $M_c$ is the mass of the companion, and $P_{\rm orb}$ is the orbital period of the binary system.

The pressure of the companion wind is
\begin{equation}
        P_{\rm cw} = \rho_{\rm cw} v_{\rm cw}^2 = \frac{\dot{M} v_{\rm cw}}{ 4 \pi  r^2},
        \label{eq:Pcw}
\end{equation}
where $\dot{M}$ is the mass-loss rate, which depends on the companion type, and $v_{\rm cw}$ is the velocity of the companion wind. Based on \cite{Vink_2000A&A...362..295V}, we estimate the mass-loss rate for O stars, B stars, and solar-like stars: 

\begin{equation}
        \label{eq:dotM}
        {\rm log} \dot{M} \simeq \left\{
        \begin{aligned}
                &-12.21, \ \ T_{c, \rm eff} \leq 12500 {\rm ~K}, \\
                &-6.688 + 2.210 ~{\rm log} \left(\frac{L_c}{10^5 \rm ~L_{\odot}}\right) \\
                &-1.339 ~{\rm log} \left(\frac{M_c}{30 \rm ~M_{\odot}}\right) - 1.601 ~{\rm log} \left(\frac{v_{\rm esc}}{v_{\rm cw, \infty}}\right) \\
                &+ 1.07 ~{\rm log} \left(\frac{T_{c, \rm eff}}{2 \times 10^4 \rm~K}\right), \\
            &\ \ 12500 {\rm ~K} < T_{c, \rm eff} \leq 22500 {\rm ~K}, \\
                &-6.697 + 2.194 ~{\rm log} \left(\frac{L_c}{10^5 \rm ~L_{\odot}}\right) - 1.313 ~{\rm log} \left(\frac{M_c}{30 \rm ~M_{\odot}}\right) \\
                &- 1.226 ~{\rm log} \left(\frac{v_{\rm esc}}{v_{\rm cw, \infty}}\right) + 0.933 ~{\rm log} \left(\frac{T_{c, \rm eff}}{4 \times 10^4 ~\rm K}\right)\\
                &- 10.92 \left[{\rm log} \left(\frac{T_{c, \rm eff}}{4 \times 10^4 \rm~K}\right)\right]^2, \\
            &\ \ 22500 {\rm ~K} < T_{c, \rm eff} \leq 100000 {\rm ~K},
        \end{aligned}
        \right.
\end{equation}
where $\dot{M}$ is in $\rm M_{\odot} ~ yr^{-1}$, $L_c$ is the luminosity of the companion, $v_{\rm cw, \infty}$ is the velocity of the stellar wind at an infinite distance from the companion, $v_{\rm esc} = (2G M_c / R_c)^{1/2}$ is the escaping velocity from the companion star, and $T_{c, \rm eff}$ is the effective temperature of the companion. We note that for solar-like stars, we set $\dot{M} \simeq 6.17 \times 10^{-13}M_\odot~{\rm yr^{-1}}$, which is consistent with the range mentioned in \cite{Wood_2002ApJ...574..412W}. We use the following mass--luminosity relation and mass--radius relation to calculate $L_c$ and $R_c$ from $M_c$ \citep{iben_2012, John_Wiley_2005}:
\begin{equation}
        \label{eq:L-M}
        L_{c, 1 L_\odot} \simeq 
        \begin{cases}
                0.23 M_{c, 1 M_{\odot}}^{2.3} & M_c \leq 0.43 {\rm~M_{\odot}} \\
                M_{c, 1 M_{\odot}}^{4} & 0.43 {\rm~M_{\odot}} < M_c \leq 2 {\rm~M_{\odot}} \\
                1.4 M_{c, 1 M_{\odot}}^{3.5} & 2 {\rm~M_{\odot}} < M_c \leq 55 {\rm~M_{\odot}} \\
                32000 M_{c, 1 M_{\odot}} & 55 {\rm~M_{\odot}}< M_c
        \end{cases}
        ,
\end{equation}
\begin{equation}
        \label{eq:R-M}
        R_{c, 1 R_\odot} \simeq 
        \begin{cases}
                M_{c, 1 M_{\odot}}^{0.8} & M_c \leq 1 {\rm~M_{\odot}} \\
                M_{c, 1 M_{\odot}}^{0.57} & 1 {\rm~M_{\odot}} < M_c
        \end{cases}
        .
\end{equation}
Furthermore, we obtain the effective temperature $T_{c, \rm eff}$ using $4 \pi R_c^2\sigma T_{c, \rm eff}^4 =  L_c$. In real observations, the effective temperature $T_{c, \rm eff}$ can be measured via the spectrum of the companion star assuming that it satisfies the black-body radiation.

Based on \cite{Vink_2000A&A...362..295V}, we use $v_{\rm esc} / v_{\rm cw, \infty} = 0.7$ for a solar-like star, $v_{\rm esc} / v_{\rm cw, \infty} = 1.3$ for a B star, and $v_{\rm esc} / v_{\rm cw,  \infty} = 2.6$ for an O star.
We can then obtain $\dot{M}$ at a given $M_c$.
In addition, the velocity of the stellar wind $v_{\rm cw}$ at $r$ is
\begin{equation}
        v_{\rm cw} = v_{\rm cw, \infty} {\left(1 - \frac{R_c}{r}\right)}^{1/2}.
        \label{eq:vcw}
\end{equation}
We approximate that $v_{\rm cw} \approx v_{\rm cw, \infty}$ when $r \gtrsim R_c$. We derive the pressure of the companion wind $P_{\rm cw}$ at a given $M_c$ and a given $r$ using Eqs.~\ref{eq:Pcw}-\ref{eq:vcw}.

\begin{figure}
        \centering
   \includegraphics[width=0.5\textwidth]{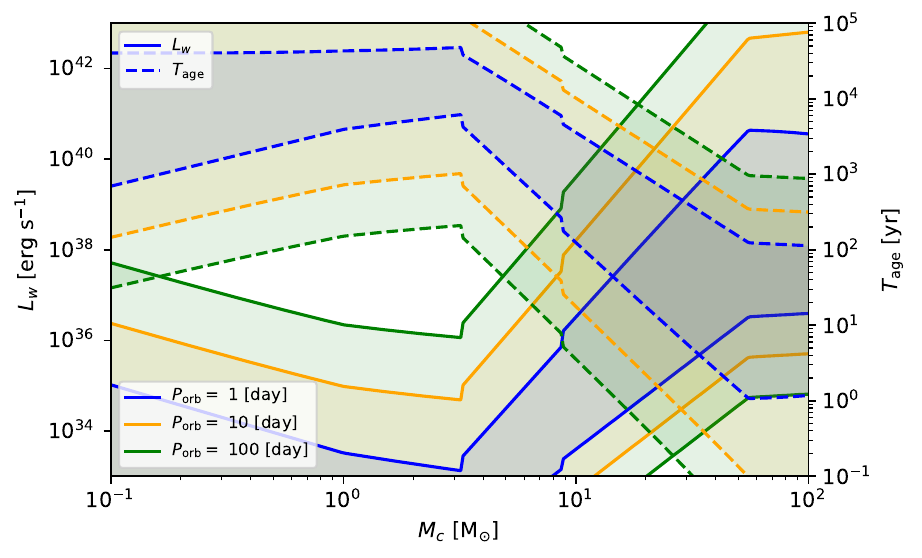}
        
        \caption{Permitted luminosity of the magnetar wind and the corresponding age of the magnetar for a given $M_c$ which varies from $0.1 ~\rm M_\odot$ to $100 ~ \rm M_{\odot}$, at which the magnetar wind will interact with the companion wind with the generation of the bow shock. Different colors represent different values of $P_{\rm orb}$, as indicated in the caption. The solid lines show the luminosity of the magnetar wind, while the dashed lines indicate the age of the magnetar. The colored areas show the permitted parameter for the scenario where the bow shock is generated from the interaction between the magnetar wind and the companion wind. }
        \label{fig:judge_bowshock}
\end{figure}

Assuming $P_w = P_{\rm cw}$ and combining with Eqs.~\ref{eq:Lw}-\ref{eq:vcw}, we can derive the location of the bow shock $r_b$ at the line connecting the two stars by numerically solving the following equation:
\begin{equation}
        \begin{aligned}
                & \frac{r_b^4}{(d-r_b)^4} = \beta_b^2 (1 - \frac{R_c}{r_b}),
        \end{aligned}
\end{equation}
where $\beta_b = L_w / \left(3 \dot{M} v_{\rm cw, \infty} c\right)$ is the ratio of the pressure for the magnetar wind and the companion wind.
When $r_b \gtrsim R_c$, we can obtain the following approximate equation:
\begin{equation}
        r_b \approx \frac{1}{1 + \sqrt{\beta_b}} a.
        \label{eq:rb_infty}
\end{equation}
According to Eq.~\ref{eq:Pcw} and Eq.~\ref{eq:vcw}, we know that $P_{\rm cw} \leq P_{\rm cw, \infty}$, where $P_{\rm cw, \infty}$ is the pressure of the companion wind with $v_{\rm cw} = v_{\rm cw, \infty}$. Therefore, we use Eq.~\ref{eq:rb_infty} to approximately determine Case I and Case II. 
Only when $r_b$ is larger than $R_c$ (corresponding to the lower limit of $T_{\rm age}$) and $d - r_b$ is larger than $R_{\rm LC}$ (corresponding to the upper limit of $T_{\rm age}$) can a bow shock be generated between a magnetar and a companion star, where $R_{\rm LC} = P_m c / (2 \pi)$ is the light cylinder of the magnetar and $P_m = P_0 \left(1 + T_{\rm age} / T_{\rm sd}\right)^{1/2}$ is the rotation period of the magnetar. 
We can then get the limitation of $L_w$ and $T_{\rm age}$ for a given $P_{\rm orb}$ and $M_c$ at which the bow shock would exist, as shown in the colored region in Fig.~\ref{fig:judge_bowshock}. We find that the larger the value of $P_{\rm orb}$, the larger the upper limit on $L_w$ and the smaller the lower limit on $T_{\rm age}$ for the bow shock.
We also find that for $P_{\rm orb} = 10 {~\rm d}$, the generation of the bow shock requires $2.6 \times 10^{30} {\rm~erg~s^{-1}} \lesssim L_w \lesssim 9.6 \times 10^{34} {\rm~erg~s^{-1}}$ ($4.8 \times 10^{32} {\rm~erg~s^{-1}} \lesssim L_w \lesssim 1.6 \times 10^{38} {\rm~erg~s^{-1}}$) with $M_c = 1 {\rm~M_{\odot}}$ ($M_c = 10 {\rm~M_{\odot}}$); 
and for $P_{\rm orb} = 100 {~\rm d}$, it requires $3.0 \times 10^{29} {\rm~erg~s^{-1}} \lesssim L_w \lesssim 2.2 \times 10^{36} {\rm~erg~s^{-1}}$ ($6.1 \times 10^{31} {\rm~erg~s^{-1}} \lesssim L_w \lesssim 4.0 \times 10^{39} {\rm~erg~s^{-1}}$) with $M_c = 1 {\rm~M_{\odot}}$ ($M_c = 10 {\rm~M_{\odot}}$).
In this way, we can divide our calculation for the persistent preburst radiation into two cases, Case I and Case II, as described above. We note that here we focus on Case I and Case II, and neglect the condition $d-r_b < R_{\rm LC}$, in which the companion wind directly interacts with the magnetar.

\subsubsection{Effects of magnetar winds on companion stars}

Now, we consider whether a magnetar has enough energy to evaporate its main sequence companion in a binary system \citep[also see][]{BHATTACHARYA19911}. The energy reservoirs of a magnetar include the rotation energy, 
\begin{equation}
    E_{\rm rot} = \frac{1}{2} I \Omega_{\rm rot}^2, 
\end{equation}
and the magnetic energy,
\begin{equation}
    E_{\rm B} \lesssim \frac{1}{6} B_{\perp}^2 R_m^3,
\end{equation}
where $P = 2 \pi / \Omega$ is the spin period of the magnetar, $B_\perp$ is the perpendicular component of the surface dipolar magnetic field of the magnetar, and $R_m$ is the radius of the magnetar. We define the efficiency $g = f (R_c / 2a)^2$ as the fraction of the magnetar's rotation energy and/or magnetic energy that actually contributes to the evaporation of the companion, where $(R_c / 2a)^2$ is the geometric factor, and $f$ refers to the "real" efficiency of the evaporation process~\citep{BHATTACHARYA19911}. The gravitational binding energy of the companion is
\begin{equation}
    E_b \simeq \frac{G M_c^2}{R_c}.
\end{equation}
The condition under which the magnetar can completely evaporate its companion is
\begin{equation}
    \frac{g \max (E_{\rm rot}, E_{\rm B})}{E_b} \geq 1.
\end{equation}
For the  following typical magnetar values, that is, $P_0 = 10^{-2} {\rm~s}$, $B_{\perp} = 10^{15} {\rm~G}$, $M_m = 1.4 {\rm~M_{\odot}}$, and $R_m = 10^6 {\rm~cm}$, a companion with a mass of $M_c = 0.1 - 20 ~M_{\odot}$ in a binary system with an orbital period of $P_{\rm orb} = 1 - 1000 ~\rm d$ cannot be completely evaporated by a magnetar even with $f = 1$.

We now consider the effects of partial evaporation on our model. The evaporation timescale of the companion $\tau_{\rm evap}$ can be estimated as
\begin{equation}
    \tau_{\rm evap} \simeq \frac{g_{\rm par} E_b}{g L_{w,0}},
\end{equation}
where $g_{\rm par} \sim m_{\rm evap} / M_c$ indicates the fraction of partial evaporation compared to complete evaporation, and $m_{\rm evap}$ is the mass of the  outer layers of the companion that may be evaporated.
If the value of $\tau_{\rm evap}$ is larger than the spin-down timescale of the magnetar $T_{\rm sd}$, the companion can survive and keep its structure, since the luminosity of the magnetar wind will drop sharply after $T_{\rm sd}$. 
For the typical values we choose for a magnetar, the evaporation timescale of a companion with $M_c = 0.1 - 20 ~M_{\odot}$ and $P_{\rm orb} = 1 - 1000 ~\rm d$ is $5.1 \times 10^5 ~{\rm s} - 2.8 \times 10^{10}~\rm s$ with $g_{\rm par} \gtrsim 0.01$ and $f = 0.01$~\citep{BHATTACHARYA19911}, which is larger than $T_{\rm sd} \approx 2.3 \times 10^5~$s.
For a smaller value of $g_{\rm par}$ ($\lesssim 0.01$), we consider the hydrostatic equilibrium timescale of the companion, which is $\tau_{\rm hyd} \simeq 3 \times 10^3 {~\rm s} \sqrt{\overline{\rho}_c / \overline{\rho}_\odot}$ (where $\overline{\rho}_\odot = 1.4 ~\rm g ~ cm^{-3}$ is the average density of the Sun, and $\overline{\rho}_c = 3 M_c / \left( 4 \pi R_c^3 \right)$ is the average density of the companion). This timescale refers to the time it takes for the companion to become spherical again after its spherical shape is destroyed. We can see that the value of $\tau_{\rm hyd} \simeq 1.3 \times 10^3 {~\rm s} - 1.5 \times 10^4 {\rm~s}$ is much smaller than $P_{\rm orb}$ and $\tau_{\rm evap}$ (unless the value of $g_{\rm par}$ is extremely small, where the evaporation process can be neglected), meaning that the companion can remain spherical in our model even if a small part of it is evaporated with $g_{\rm par} \lesssim 0.01$. 

Therefore, we believe that, at least within the confines of the model we study, the companion star can remain spherical and its evaporation process can be neglected.
Below, for simplicity, we use the Lane-Emden model with $n = 3$ to simulate the internal structure of the companion star. 
This model is associated with a fully radiative star, where it is assumed that the ratio between the gas pressure and the total pressure is a constant throughout the star, which fits well with the internal structure of the Sun \citep{Guenther_1992ApJ...387..372G} and very massive stars \citep{Nadyozhin_2005AstL...31..695N}. 
The Lane-Emden equation is
\begin{equation}
        \frac{1}{\xi^2} \frac{d}{d \xi} \left(\xi^2 \frac{d \theta_{\rm LE}}{d \xi}\right) = - \theta_{\rm LE}^n,
        \label{eq:LE}
\end{equation}
where $\theta_{\rm LE}$ and $\xi = r / r_n$ is the dimensionless parameter, $r_n = {\left[ (n+1) K \rho_{\rm ce}^{1/n-1} / (4 \pi G) \right]}^{1/2}$ is a typical radius, $K$ is a constant, and $\rho_{\rm ce} = M_c / \left[ 4 \pi R_c^3 {\left( - \frac{1}{\xi} \frac{d\theta_{\rm LE}}{d\xi}  \right)}_{\xi = \xi_1}  \right]$ is the density at the center of the companion star. We note that $\xi_1 = R_c / r_n$ indicates the surface of the companion star, where $\theta_{\rm LE} (\xi_1) = 0$. The internal density of the companion is $\rho_c = \rho_{\rm ce} \theta_{\rm LE}^n$, and the internal pressure of the companion is $P_c = K \rho_c^{1 + \frac{1}{n}}$. 
Numerically solving Eq.~\ref{eq:LE}, we can obtain the internal structure of the companion for a given $M_c$, including $P_c$ and $\rho_c$ at each radius $r = \xi r_n$.

\subsubsection{Radiation from the direct interaction between magnetar wind and companion star}
\label{sect:direct}

In Case I, the luminosity of the magnetar wind is sufficiently high that it will directly interact with the companion star and heat it. 
Following \cite{Yang_2021ApJ...920...34Y}, who analyzed the process by which an FRB heats the companion surface, we consider that the magnetar wind terminates at $r_b$. The value of $r_b$ is determined by numerically solving the equation $P_w (r) = P_c (r)$ using Eq.~\ref{eq:Pw} and Eq.~\ref{eq:LE}. The thickness of the shocked medium at $\theta_m = 0$ is $l = R_c - r_b$. 
We note that for $M_c = 0.1 - 20 ~M_\odot$ and $P_{\rm orb} = 1 - 1000 \rm~d$, the value of $l / R_c$ is smaller than $0.1,$ meaning that the spherical shape of the companion star is not significantly affected, and we can still use the Lane-Emden equation to approximate the structure of the companion.

We construct a spherical coordinate system with the center of the magnetar as the origin and the vector from the magnetar to the companion star as the $z$-axis. We divide the heating region into two parts, $\theta_m \in (0, \theta_{m1})$ (region I) and $\theta_m \in (\theta_{m1}, \theta_{m2})$ (region II), where $\theta_{m1} = \arcsin \frac{r_b}{a}$ and $\theta_{m2} = \arcsin \frac{R_c}{a}$, as shown in Fig.~\ref{fig:windmodel1}. The volume of the heating region of the companion star is
\begin{equation}
        \begin{aligned}
                & V_{\rm sw} = 2 \pi \int_0^{\theta_{m1}} \int_{r'_{x_1}}^{r'_{y_1}} r'^2 \sin \theta_m dr' d\theta_m \\
                & + 2 \pi \int_{\theta_{m1}}^{\theta_{m2}} \int_{r'_{x_2}}^{r'_{y_2}} r'^2 \sin \theta_m dr' d\theta_m,
        \end{aligned} 
\end{equation}
where $r'_{x_1} = r'_{x_2} = a \cos \theta_m - \sqrt{a^2 (\cos^2 \theta_m - 1) + R_c^2}$ and $r'_{y_1} = a \cos \theta_m - \sqrt{a^2 (\cos^2 \theta_m - 1) + r_b^2}$ are the boundary of region I, and $r'_{x_1} = r'_{x_2}$ and $r'_{y_2} = a \cos \theta_m + \sqrt{a^2 (\cos^2 \theta_m - 1) + R_c^2}$ are the boundary of region II. 
The mass of the heating region is
\begin{equation}
        \begin{aligned}
                & m_{\rm sw} = 2 \pi \int_0^{\theta_{m1}} \int_{r'_{x_1}}^{r'_{y_1}} \rho_c r'^2 \sin \theta_m dr' d\theta_m \\
                & + 2 \pi \int_{\theta_{m1}}^{\theta_{m2}} \int_{r'_{x_2}}^{r'_{y_2}} \rho_c r'^2 \sin \theta_m dr' d\theta_m.
        \end{aligned} 
        \label{eq:msw}
\end{equation}
We note that the distance between the point in the heating region and the center of the companion is $r_l = \sqrt{r'^2 - 2dr \cos \theta_m + a^2}$, and we can then derive the density $\rho_c = \rho_c (r_l)$ for a given $r'$. In this way, the mass of the heating region can be calculated.
The injection energy is
\begin{equation}
        \begin{aligned}
                & \mathcal{E}= 2 \pi \int_0^{\theta_{m1}} 
                \int_{r'_{x_1}}^{r'_{y_1}} \frac{L_w}{4 \pi r'^2 c} r'^2 \sin \theta_m dr d\theta_m \\
                & + 2 \pi \int_{\theta_{m1}}^{\theta_{m2}} \int_{r'_{x_2}}^{r'_{y_2}} \frac{L_w}{4 \pi r'^2 c} r'^2 \sin \theta_m dr' d\theta_m \\
                & = \frac{L_w}{2 c} \int_0^{\theta_{m1}} 
                \int_{r'_{x_1}}^{r'_{y_1}} \sin \theta_m dr d\theta_m + \frac{L_w}{2 c} \int_{\theta_{m1}}^{\theta_{m2}} \int_{r'_{x_2}}^{r'_{y_2}} \sin \theta_m dr' d\theta_m.
        \end{aligned} 
        \label{eq:E_direct}
\end{equation}
This energy will transfer to particles of the heating region. The number of these particles and the temperature of the heating region should be~\citep{Yang_2021ApJ...920...34Y}
\begin{equation}
        \begin{aligned}
                & N = \frac{f_b m_{\rm sw}}{m_p}, \\
                & T = \frac{\eta \mathcal{E}}{N k},
        \end{aligned}
        \label{eq:NandT}
\end{equation}
where $\eta \sim 0.01 - 0.1$ is the factor accounting for energy losses during the energy transformation process, and $f_b \sim 0.1 - 1$ is a corrected fraction for the mass of the companion swept by the magnetar wind since the above mass is an overestimated approximation. 
Here, we use $\eta = 0.01$ and $f_b = 0.1$ for an example.

Considering the radiative transfer process, the optical depth for the Thomson scattering at $\theta_m = 0$ can be estimated as 
\begin{equation}
        \tau \simeq \int_{0}^{l}   \kappa \rho_c(r_b + l_r) dl_r, 
        \label{eq:tau}
\end{equation}
where $\kappa \sim 0.4 {~\rm cm^2 ~g^{-1}}$ is the Thomson scattering opacity of fully ionized hydrogen, and $\rho_c(r_b + l_r)$ is the internal density of the companion at $r_b + l_r$ ($l_r$ is from $0$ to $l$). We use this optical depth to approximately calculate the effective temperature of the heating region based on the theory of radiative transfer~\citep{Yang_2021ApJ...920...34Y}:
\begin{equation}
        T_{\rm eff} = T {\left( \frac{1}{2} + \frac{3}{4} \tau \right)}^{-1/4}.
        \label{eq:Teff}
\end{equation}
Based on the above equations and the black-body radiation theory, the total luminosity of re-emission generated by the magnetar wind heating the companion can be estimated as
\begin{equation}
        L_{\rm re} \simeq 2 \pi R_c^2 \sigma_{\rm SB} T_{\rm eff}^4,
        \label{eq:Lre}
\end{equation}
where $\sigma_{\rm SB}$ is the Stefan-Boltzmann constant. The specific luminosity is 
\begin{equation}
        L_{\nu} = 2 \pi^2 R_c^2 \frac{2 h \nu^3}{c^2} \frac{1}{e^{h \nu / (k T_{\rm eff})} - 1}.
        \label{eq:Lnu}
\end{equation}

In addition, we compare the gravitational binding energy of the swept mass of the companion $g_{\rm par} E_b \simeq G m_{\rm sw} (M_c - m_{\rm sw}) / r_b$ and the injection energy $\eta \mathcal{E}$. We find that in typical parameter spaces we choose, the value of $\eta \mathcal{E}$ is much smaller than $g_{\rm par} E_b$, and their ratio is $\lesssim 10^{-3}$. Therefore, we neglect the evaporation process of the companion caused by the magnetar wind in our work.

\subsubsection{Radiation from the interaction between magnetar wind and companion wind}
\label{sect:bowshock}

In Case II, the interaction between the magnetar wind and the companion wind generates a bow shock at $r_{\rm b}$ when $L_w$ is small enough, as shown in Fig.~\ref{fig:judge_bowshock}. 
Following the analytical calculations in \cite{Wadiasingh_2017ApJ...839...80W} and \cite{Kong_2011MNRAS.416.1067K}, we calculate the synchrotron radiation from the bow shock. We use the geometry of the bow shock proposed in \cite{Cant_1996ApJ...469..729C}. 
For more complex cases, a hydrodynamic simulation was performed to study this issue~\citep{Zabalza_2013A&A...551A..17Z} and two-region models of two shocks have been studied~\citep{Usov_1992ApJ...389..635U}. It is noteworthy that the structure of the bow shock calculated by the analytical method  is approximately consistent with that found using a numerical method \citep{Zabalza_2013A&A...551A..17Z}. 
Also, many previous studies~\citep[e.g.,][]{Wadiasingh_2017ApJ...839...80W,Chen_2019A&A...627A..87C,Chen_2021A&A...652A..39C} that use similar methods to that presented here can model the observations of gamma-ray binaries and millisecond pulsar binaries very well. In this way, as part of our analysis, we use the above model to calculate the radiation. 
For $\beta_b > 1$, the magnetar wind is stronger, meaning that the bow shock bends toward the companion. The shape of the bow shock using the linear and angular momentum conservation is
\begin{equation}
        \begin{aligned}
                & r_b (\theta) = d \sin \theta_m \csc (\theta + \theta_m), \\
                & r_s (\theta) = d \sin \theta \csc (\theta + \theta_m), \\
                & \theta_m \cot \theta_m = 1 + \beta_b (\theta \cot \theta - 1),
        \end{aligned}
        \label{eq:bowshock_shape}
\end{equation}
where $\theta$ is the angle measured from the line between two stars with the companion as the origin, $\theta_m$ is the angle with the magnetar as the origin, and $r_s$ is the distance between the bow shock and the magnetar, as shown in Fig.~\ref{fig:windmodel2}. When $\theta = 0$, the location of the bow shock can be derived using Eq.~\ref{eq:rb_infty}. The asymmetric angle $\theta_\infty$ of the bow shock wing, corresponding to $r_b \to \infty,$ can be calculated using 
\begin{equation}
        \theta_\infty - \tan \theta_\infty = \frac{\pi}{1 - \beta_b}.
\end{equation}
We estimate the volume of the radiation region to be 
\begin{equation}
        \begin{aligned}
                V_{\rm bow} = 2 \pi \Delta_{\rm bow} \int_0^{\theta_\infty} \int_0^{r_b (\theta)} r^2 \sin \theta d r d \theta,
        \end{aligned}
        \label{eq:Vbow}
\end{equation}
where $\Delta_{\rm bow} = \frac{1}{8} {\rm min} \left[r_b (\theta=0), r_s (\theta=0)\right]$ is the estimated thickness of the shocked shell \citep{Luo_1990ApJ...362..267L}. Conversely, for $\beta_b < 1$, the companion wind is stronger, and the bow shock bends toward the magnetar. For Eqs.~\ref{eq:bowshock_shape}-~\ref{eq:Vbow}, we need to replace $r_b$ with $r_s$ and $\theta_m$ with $\theta$ for $\beta_b < 1$. 
We estimate the equatorial upstream wind magnetic field to be~\citep[e.g.,][]{Kong_2011MNRAS.416.1067K}
\begin{equation}
        \begin{aligned}
                & B_w^2 =  \frac{\sigma L_w}{r_s^2 c (1 + \sigma)}.
        \end{aligned}
        \label{eq:B_w}
\end{equation}
We note that for different $\theta$ and/or $\theta_m$, we have different values of $B_w$.
\cite{Kennel_1984ApJ...283..694K} and \cite{Kennel_1984ApJ...283..710K} suggested that the pulsar wind magnetization $\sigma$, that is, the ratio of the magnetic energy flux to particle energy flux upstream of the shock, is $\sim 0.003$ for the Crab pulsar wind nebula (PWN) based on observations, while \cite{Shibata_2003MNRAS.346..841S} proposed a modified model considering magnetic-field turbulence, which gives $\sigma \sim 0.05$. Here we use $\sigma = B_w^2 / (4 \pi n_{e,w} \left<\gamma_w\right> m_e c^2) = 0.01$ to obtain an estimation of the mean Lorentz factor of the radiation region $\left<\gamma_w\right>$~\citep{Wadiasingh_2017ApJ...839...80W,van_2020ApJ...904...91V}:
\begin{equation}
        \left<\gamma_w\right> \simeq {\left(\frac{L_w}{c}\right)}^{1/2} \frac{5e}{\mathcal{M}_{\pm} m_e c^2},
        \label{eq:gamma_w}
\end{equation}
where $e$ is the electron charge, $n_{e,w} \approx \mathcal{M}_{\pm} \sqrt{6 c L_w} / \left(4 \pi e c r_s^2\right)$ is the total number density of electrons in the radiation region contributed by the magnetar wind, and $\mathcal{M}_{\pm}$ is the pair multiplicity of the Goldreich-Julian rate. 
The value of $\mathcal{M}_{\pm}$ for a magnetar could be very different, because it highly depends on parameters (such as $P$ and $B_{\perp}$) due to the photon splitting and pair creation~\citep[e.g.,][]{Thompson_2008ApJ...688..499T,Chen_2017ApJ...844..133C}.
The theoretical estimation of multiplicity for the magnetar is $\sim 10^2 - 10^4$~\citep{Medin_2010MNRAS.406.1379M,Beloborodov_2013ApJ...762...13B}, while the young pulsar wind nebula and double pulsar studies~\citep[e.g.,][]{Sefako_2003ApJ...593.1013S,Breton_2012ApJ...747...89B} suggest $\mathcal{M}_{\pm} \sim 10^3 -10^5$~\citep{Wadiasingh_2017ApJ...839...80W}.
Here we use $\mathcal{M}_{\pm} = 10^4$ as the typical value.
Based on the shock jump condition for small $\sigma$ in \cite{Kennel_1984ApJ...283..710K}, the magnetic field of the shocked region is $B_{\rm bow} = 3 (1 - 4 \sigma) B_w \simeq 3 B_w$, and the Lorentz factor of the shocked region is $\gamma_f \approx \sqrt{(9 + 9 \sigma) / 8} \simeq 1,$ which means that the Doppler effect is not important. Based on the results in \cite{Wadiasingh_2017ApJ...839...80W}, the orbital changes with the period and the direction of the line of sight affect the luminosity by about one order of magnitude at most. 
We ignore the influence of these factors here, because our studies (as shown in Sect.~\ref{sect:lcandsp} and Sect.~\ref{sect:param}) find that the luminosity of a bow shock is relatively weak.

We assume the differential electron number density distribution in the fast-cooling case ($\gamma_m > \gamma_c$) is
\begin{equation}
        \begin{aligned}
                N_e (\gamma, \theta) d \theta & = \frac{d N_e d \theta}{d \gamma} \\
                & = N_{e,0} (\theta) d \theta {\left( \frac{1}{\gamma_c} - \frac{1}{\gamma_m} + \frac{\gamma_m^{-p} - \gamma_M^{-p}}{p} \gamma_m^{p-1} \right)}^{-1} \\
                & \times
                \begin{cases}
                        \gamma^{-2} & \gamma_c \leq \gamma < \gamma_m \\
                        \gamma_m^{p-1} \gamma^{-(p+1)} & \gamma_m \leq \gamma < \gamma_M,
                \end{cases}
        \end{aligned}
\end{equation}
and that in the slow-cooling case ($\gamma_m < \gamma_c$) it should be
\begin{equation}
        \begin{aligned}
                N_e (\gamma, \theta) d \theta & = N_{e,0} (\theta) d \theta {\left( \frac{\gamma_c^{-p+1} - \gamma_m^{-p+1}}{1-p} + \frac{\gamma_c^{-p} - \gamma_M^{-p}}{p} \gamma_c \right)}^{-1} \\
                & \times
                \begin{cases}
                        \gamma^{-p} & \gamma_m \leq \gamma < \gamma_c \\
                        \gamma_c \gamma^{-(p+1)} & \gamma_c \leq \gamma < \gamma_M,
                \end{cases}
        \end{aligned}
\end{equation}
where $N_{e,0} (\theta) d \theta = n_{e,w} \Delta_{\rm bow} r_b^2 \sin \theta d\theta$ is the number of electrons in the radiation region at a given $\theta$, $\gamma_m \simeq \left<\gamma_w\right> (p-2) / (p-1)$ and $\gamma_M \simeq \sqrt{6 \pi e / \left( \sigma_T B_{\rm bow} \right)}$ are the minimum and maximum Lorentz factor of accelerated electrons, $\sigma_T$ is the Thompson scattering cross-section, $\gamma_c \simeq 6 \pi m_e c / \left(\sigma_T t_{\rm dyn} B_{\rm bow}^2 \right)$ is the synchrotron cooling Lorentz factor, and $t_{\rm dyn} \simeq \xi_f R_{\rm bow} / v_f$ with $R_{\rm bow} = \min (r_b, r_s)$ is the estimated dynamic flow time of electrons in the radiation region \citep{Kong_2011MNRAS.416.1067K}. Based on \cite{Tavani_1997ApJ...477..439T} and \cite{Kennel_1984ApJ...283..710K}, we use $v_f = c / 3,$ which can be derived from the shock jump condition for a small $\sigma$, and we use the coefficient $\xi_f = 3,$ considering the nonspherical shape of the shocked region. We note that when $\gamma_c > \gamma_M$, the differential electron number density distribution is
\begin{equation}
        \begin{aligned}
                N_e (\gamma, \theta) d \theta & = N_{e,0} (\theta) d \theta \frac{1-p}{(\gamma_M^{-p+1} - \gamma_m^{-p+1})} 
                \gamma^{-p}, ~~~ \gamma_m \leq \gamma < \gamma_M.
        \end{aligned}
\end{equation}
The synchrotron luminosity in the observer frame should be \cite[e.g.,][]{Rybicki_1985rpa..book.....R}
\begin{equation}
        \begin{aligned}
                & L_{\rm bow} = \nu \int_0^{\theta_\infty} \int_{\gamma_n}^{\gamma_M} P(\gamma, \nu, \theta) N_e(\gamma, \theta) d\gamma d \theta ~~~{\rm with} \\
                & P(\gamma, \nu, \theta) = \frac{\sqrt{3} e^3 B_{\rm bow}}{m_e c^2} F({\nu} / {\nu_{\rm ch}}), \\
                & F({\nu} / {\nu_{\rm ch}}) \equiv x \int^{\infty}_x K_{5/3} (\zeta) d\zeta, \\
        \end{aligned}
        \label{eq:L_bow}
\end{equation}
where $P(\gamma, \nu)$ is the emitted power per unit frequency, $\nu_{\rm ch} = 3 \gamma^2 e B_{\rm bow} / ( 4 \pi m_e c)$ is the characteristic emission frequency, and $\gamma_n = {\rm min} ~(\gamma_m, \gamma_c)$. We use $L_{\rm re} \sim L_{\rm bow}^{\rm max}$ to approximate the total luminosity, where $L_{\rm bow}^{\rm max}$ is the peak value of $L_{\rm bow}$.

\subsection{Radiation from bursts heating the companion star}
\label{sect:burst}

In addition to the magnetar wind, the magnetar can generate different kinds of bursts during the active phases. We consider three kinds of bursts: MSBs with $\Delta t_{\rm burst} \sim 0.1 {~\rm s}$ and $\mathcal{E}_{\rm burst} \sim 10^{41} {~\rm erg}$ \citep{Mereghetti_2008A&ARv..15..225M}, MGFs with $\Delta t_{\rm burst} \sim 0.1 {~\rm s}$ and $\mathcal{E}_{\rm burst} \sim 10^{46} {~\rm erg}$ \citep{Burns_2021ApJ...907L..28B}, and FRBs with $\Delta t_{\rm burst} \sim 10^{-3} {~\rm s}$ and $\mathcal{E}_{\rm burst} \sim 10^{39} {~\rm erg}$ \citep{Luo_2020MNRAS.494..665L}. Here we take the typical value of these bursts, and $\Delta t_{\rm burst}$ is the duration of bursts and $\mathcal{E}_{\rm burst}$ is the isotropic energy of bursts. 
We note that MGFs are expected to involve baryonic matter in their outflows~\citep[e.g.,][]{Granot_2006ApJ...638..391G}. However, in our model, we consider that bursts heat the surface of the companion, where the released energy is transformed into thermal energy, leading to the emission of black-body radiation. Therefore, considering the thermalized process on the companion's surface, the heating process of MGF is similar to the other types of bursts, except for their greater energy.

\subsubsection{Effects of bow shock on bursts}

We set the typical frequency for FRBs to be $\nu_{\rm burst} = 1~\rm GHz$, and that for MSBs and MGFs to be $\nu_{\rm burst} = 2.4 \times 10^{17} {\rm~Hz}$ corresponding to $1~{\rm keV}$. If the age of the magnetar is relatively old, due to the weaker magnetar wind, a bow shock will be generated instead of the direct interaction with the companion surface. We are interested in whether the bursts could be absorbed by the bow shock medium via the synchrotron self-absorption (SSA) process. The optical depth of SSA contributed by the relativistic electrons in the bow shock region is
\begin{equation}
        \begin{aligned}
                & \tau_{\rm ssa} \simeq \alpha_{\rm ssa} \Delta_{\rm bow}~~~{\rm with} \\
                & \alpha_{\rm ssa} = -\frac{1}{8 \pi \nu^2 m_e}  \int_{\gamma_n}^{\gamma_M} P(\gamma, \nu) \gamma^2 \frac{\partial}{\partial \gamma} \left[\frac{N_e(\gamma)}{\gamma^2}\right] d\gamma.
        \end{aligned}
\end{equation}
We find that FRBs, MSBs, and MGFs all have $\tau_{\rm ssa} \gg 1$. However, we note that if the internal pressure of the bow shock is less than the pressure of the burst, the bursts will break through the bow shock and still directly heat the companion surface, even if $\tau_{\rm ssa} \gg 1$. Based on the jump conditions for small $\sigma,$ assuming the upstream is highly relativistic \citep{Kennel_1984ApJ...283..710K}, the pressure of the bow shock region can be estimated to be
\begin{equation}
        P_{\rm bow} \simeq \frac{2}{3} (1 - 7 \sigma) n_{e,w} m_e c^2 \left<\gamma_w\right>,
\end{equation} 
and the radiation pressure of the burst is
\begin{equation}
        P_{\rm burst} \simeq \frac{\mathcal{E}_{\rm burst}}{4 \pi c (d - r)^2 \Delta t_{\rm burst}}.
\end{equation}
We find that FRBs, MSBs,and MGFs all have $P_{\rm burst} \gg P_{\rm bow}$ for the typical parameters chosen here, implying that the bursts will destroy the bow shock and then interact with the companion. 
Therefore, similar to the model of Case I, we consider the burst will directly interact with the companion star even if the bow shock is generated, and the energy of the burst will be transferred to the swept medium, generating the re-emission with a re-emission timescale of $\Delta t_{\rm re}$. 

\subsubsection{Absorption processes when bursts heat the companion}

Here we discuss how a burst from the magnetar heats the companion star. 
We consider two types of absorption processes for the process of a burst interacting with the companion, including plasma absorption and free-free (FF) absorption. 
The plasma frequency of the  medium on the surface of the  companion is \citep{Tonks_1929PhRv...33..195T}
\begin{equation}
        \nu_{\rm pe} = \left( \frac{n_{e, c} e^2}{\pi m_e} \right)^{1/2},
\end{equation}
where $n_{e, c} \simeq \rho_c / m_p$ is the number density of the electrons in the companion. The photons in bursts with $\nu_{\rm burst} > \nu_{\rm pe}$ can penetrate the plasma. We find that for FRBs, they are almost impenetrable. For MSBs and MGFs, they can penetrate into very inner layers $r_{\rm ph} \lesssim 0.01 R_c$. For the FF absorption, we use \citep{Lang_1999acfp.book.....L,Murase_2017ApJ...836L...6M}
\begin{equation}
        \begin{aligned}
                & \tau_{\rm ff} \simeq \int_{0}^{R_c - r_{\rm ph}} \alpha_{\rm ff} d l_r ~~~{\rm with}\\ 
                & \alpha_{\rm ff} \simeq 8.5 \times 10^{-28} \ \bar{Z}^2 {\left(\frac{\nu}{10^{10} \ \rm Hz}\right)}^{-2.1} n_{e, c} n_{i, c} \\
                & \times {\left(\frac{T_{c, \rm eff}}{10^4 \ \rm K}\right)}^{-1.35} {\left(\frac{1-e^{-h\nu / kT_{\rm eff, c}}}{h \nu / k T_{\rm ext}}\right)},
        \end{aligned}
\end{equation}
where $\bar{Z} = 1$ is the charge number assuming that hydrogen is dominated, $T_{c, \rm eff}$ is the effective temperature of the companion, and $n_{i, c}$ is the ion number density in the companion. We take $n_{i, c} = n_{e, c}$ for the assumption that hydrogen is dominated at the companion envelope. Combining with the internal structure of the companion, we can then obtain the location of the photosphere for the FF absorption using $\tau_{\rm ff} = 1$. We find that for FRBs, MSBs, and MGFs, the location of the photosphere is always on the surface of the companion star, $r_{\rm ph} \approx R_c$. Considering both absorption processes, FRBs, MSBs, and MGFs are all absorbed at the surface of the companion star. However, due to dynamic compression, the shock generated from the bursts interacting with the companion will still sweep the companion star and stop at $P_{\rm burst} = P_c$.

\subsubsection{Re-emission generated by bursts heating the companion}

Based on the above calculations, similar to the calculation in Sect.~\ref{sect:direct}, assuming $\Delta t_{\rm burst} \ll \Delta t_{\rm re}$, the injection energy should be 
\begin{equation}
        \mathcal{E}= \left(\frac{\Delta \Omega}{4 \pi} \right) \mathcal{E}_{\rm burst},
        \label{eq:E_burst}
\end{equation}
where $\Delta \Omega \sim \pi R_c^2 / d^2$ is the solid angle of the companion star opened to the magnetar. We assume that the solid angle of bursts $\Delta \Omega_{\rm burst}$ is larger than $\Delta \Omega$ for the three kinds of burst considered here.
We can also calculate $r_b$ by numerically solving $P_{\rm burst} (r) = P_c (r)$.
Combining with Eqs.\ref{eq:msw}-\ref{eq:Lnu} and substituting Eq.~\ref{eq:E_burst} for Eq.~\ref{eq:E_direct}, we can then derive the re-emission luminosity for this process. Unlike the radiation from the magnetar wind heating the companion, this process is a transient process with the re-emission timescale $\Delta t_{\rm re}$:
\begin{equation}
        \Delta t_{\rm re} \simeq \frac{\eta \mathcal{E}}{L_{\rm re}}.
\end{equation}
The temperature evolution of the heating region due to radiation cooling is~\citep{Yang_2021ApJ...920...34Y}
\begin{equation}
        t = \frac{N k}{6 \pi R_c^2 \sigma} \left(\frac{1}{T_{\rm eff}} - \frac{1}{T_{\rm eff,0}^3}\right),
        \label{eq:lc}
\end{equation}
where $T_{\rm eff, 0}$ is the initial temperature that we can obtain from Eq.~\ref{eq:Teff}. We can then use Eq.~\ref{eq:Lre} and Eq.~\ref{eq:Lnu} to obtain the light curve of the re-emission, because the re-emission is the black-body radiation.

However, in our calculation, we find in some cases that $\Delta t_{\rm re} \ll \Delta t_{\rm burst}$ where the process cannot be thought of as an instantaneous injection of energy, but as a stable process. Therefore, in these cases, we still have to use Eq.~\ref{eq:E_direct} to calculate the injection energy $\mathcal{E}$ by substituting $L_{\rm burst} = \mathcal{E}_{\rm burst} / \Delta t_{\rm burst}$ for $L_w$. We note that when calculating $\mathcal{E}$, we need to add an extra limitation, which is $r'_y - r'_x \leq c \Delta t_{\rm burst}$ for each $\theta,$ because this is still a short process. When $r'_y - r'_x > c \Delta t_{\rm burst}$, $r'_y$ should be $r'_x + c \Delta t_{\rm burst}$. For the calculation of the light curve, we assume that $L_{\rm re}$ is a constant when $t < \Delta t_{\rm burst}$. When $t \gtrsim \Delta t_{\rm burst}$, we still use Eq.~\ref{eq:lc} to calculate the cooling process.

In addition, we compare the gravitational binding energy of the swept mass of the companion $g_{\rm par} E_b \simeq G m_{\rm sw} (M_c - m_{\rm sw}) / r_b$ and the injection energy of the bursts $\eta \mathcal{E}$. We find that in the typical parameter spaces we choose, the value of $\eta \mathcal{E}$ is much smaller than $g_{\rm par} E_b$, and their ratio is $\lesssim 10^{-2}$. Therefore, we neglect the evaporation process of the outer layers of the companion caused by magnetar bursts in our work.

\section{Results}

\subsection{Spectra and light curves}
\label{sect:lcandsp}

We show the spectra and light curves with $P_{\rm orb} = 1 {\rm~d}$ for a companion with $M_c = 1 {\rm~M_{\odot}}$ and $10 {\rm~M_{\odot}}$. Both kinds of persistent preburst radiation are considered. For Case I (blue lines), we set $L_w \approx 9.6 \times 10^{44} {\rm~erg~s^{-1}}$, corresponding to a newborn magnetar in which the magnetar wind would directly interact with the companion. For Case II (light blue lines), we set $L_w \approx 5.0 \times 10^{34} {\rm~erg~s^{-1}}$ corresponding to a magnetar with $T_{\rm age} = 10^3 {\rm~yr}$ in which the bow shock will be generated.

\begin{figure*}
        \centering
        \subfigure[]{
                \includegraphics[scale=0.6]{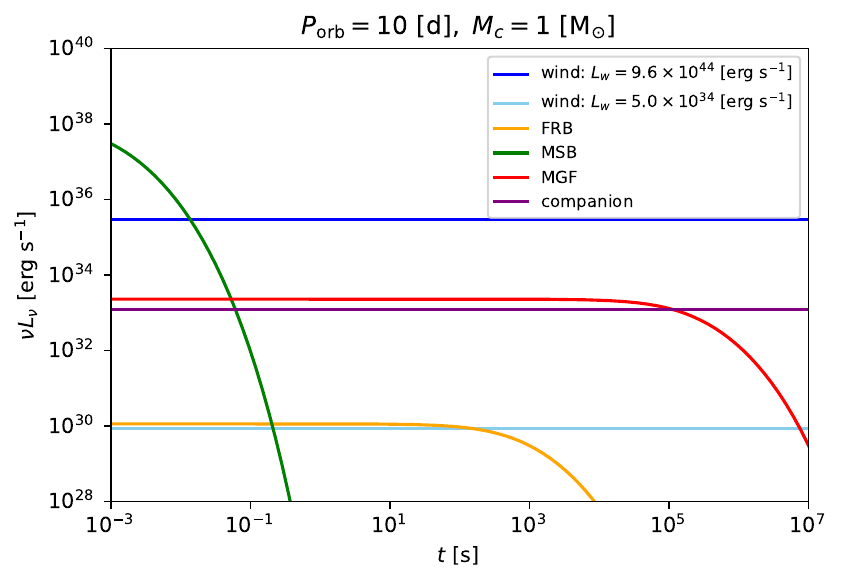}
                \label{fig:lc1}
        }%
        \subfigure[]{
                \includegraphics[scale=0.6]{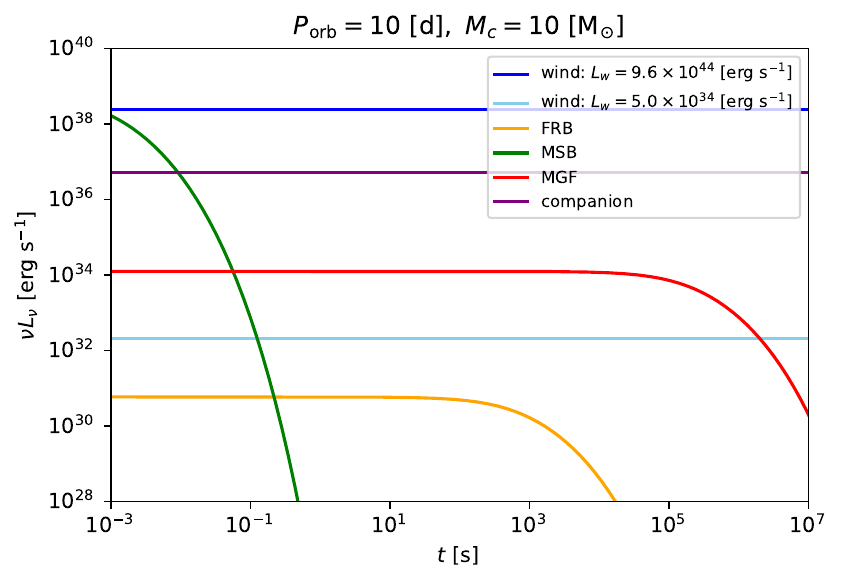}
                \label{fig:lc2}
        }%
        
        \subfigure[]{
                \includegraphics[scale=0.6]{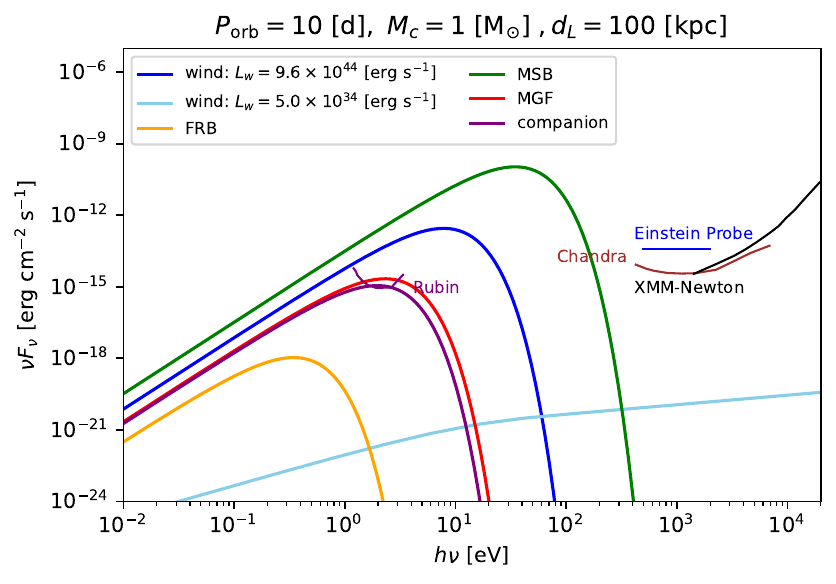}
                \label{fig:sp1}
        }%
        \subfigure[]{
                \includegraphics[scale=0.6]{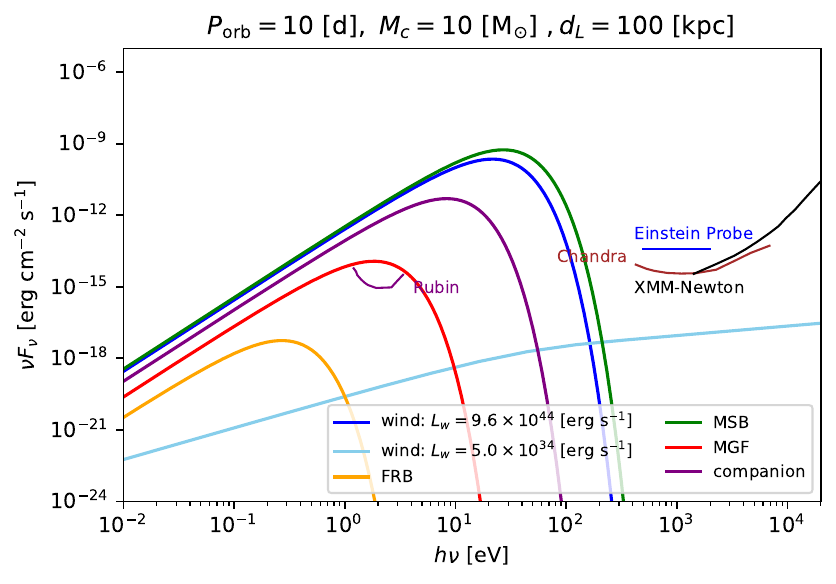}
                \label{fig:sp2}
        }%
        \caption{Spectra and light curves with $M_c = 1 {\rm~M_{\odot}}$ (left panels) and $10 {\rm~M_{\odot}}$ (right panels) in a binary system with $P_{\rm orb} = 10 {\rm~d}$. Differently colored lines represent different mechanisms, as indicated in the legends. The blue line shows the persistent preburst radiation for $L_w = 9.6 \times 10^{44} {\rm~erg~s^{-1}}$ corresponding to Case I, and the light blue line indicates that for $L_w = 5.0 \times 10^{34} {\rm~erg~s^{-1}}$ corresponding to Case II. In the top panels, we show the light curves. In the bottom panels, we show the spectra for different mechanisms at a distance of $d = 100 {\rm~kpc}$. The sensitivity curves of \textit{Einstein Probe} (blue line), \textit{Chandra} (pink line), and \textit{XMM-Newton} (black line) are calculated with an exposure time of $10^3 {\rm~s,}$ which can be derived from sensitivity proportional to the $-1/2$ power of the exposure time \citep{Lucchetta_2022JCAP,Yuan_2022hxga.book...86Y}, and the r-band sensitivity of the Vera C. Rubin Observatory (purple line) is calculated with a point source exposure time of $30 {\rm~s}$ \citep{Yuan_2021ApJ...911L..15Y}.}
        \label{fig:lsandsp}
\end{figure*}

In Fig.~\ref{fig:lc1} and Fig.~\ref{fig:lc2}, we show the light curves of the persistent preburst radiation and the re-emission from bursts with $M_c = 1 {\rm~M_{\odot}}$ and $10 {\rm~M_{\odot}}$, where we use the peak luminosity from the spectra to estimate the total luminosity for each radiation mechanism. 
We can see that the peak luminosity of persistent preburst radiation with $L_w \approx 9.6 \times 10^{44} {\rm~erg~s^{-1}}$ is dominant all the time except for the very early time of $\lesssim 0.01~$s, where the re-emission from MSBs is dominant, while that with $L_w \approx 5.0 \times 10^{34} {\rm~erg~s^{-1}}$ is outshined by the radiation from the companion itself all the time. 
In Fig.~\ref{fig:lc1} with $M_c = 1 {\rm~M_{\odot}}$, we can see that the re-emission from MGFs (MSBs) is greater than the luminosity of the companion with $t \lesssim 1.2 \times 10^5 {\rm~s}$ ($0.06 {\rm~s}$), and the re-emission from FRBs is outshined by $L_c$ all the time. 
In Fig.~\ref{fig:lc2} with $M_c = 10 {\rm~M_{\odot}}$, the re-emission from MSBs is larger than $L_c$ with $t \lesssim 0.09 {\rm~s}$, and that from FRBs and MGFs is outshined by $L_c$ all the time.
We note that the re-emission from FRBs still has the possibility of being observed due to the uncertainty of its transformation efficiency.

In Fig.~\ref{fig:sp1} and Fig.~\ref{fig:sp2}, we show the spectra at $d_L = 100 {\rm~kpc}$ with $M_c = 1 {\rm~M_{\odot}}$ and $10 {\rm~M_{\odot}}$ for $P_{\rm orb} = 10 {\rm~d}$, where $d_L$ is the distance between the source and the observer. The persistent preburst radiation for Case I is mainly at the optical and ultraviolet (UV) bands, and that for Case II is mainly at the X-ray band and higher energy band. The re-emission from bursts is mainly at the optical, UV, and X-ray bands.
We note that the re-emission from MSBs is a short process with $\Delta t_{\rm re} \sim 0.1$ in the parameter we choose, while the exposure times for the sensitivity curves shown in Fig.~\ref{fig:sp1} and Fig.~\ref{fig:sp2} are $30 {\rm~s}$ and $10^3 {\rm~s}$. The sensitivity for detectors with shorter exposure time would be lower, meaning that the re-emission from MSBs would be difficult to detect.
For the optical band, the Vera C. Rubin Observatory would be capable of detecting the persistent preburst radiation with $L_w \approx 9.6 \times 10^{44} {\rm~erg~s^{-1}}$ (for $M_c = 1 \rm ~M_{\odot}$ and $10 \rm ~M_{\odot}$) and the re-emission from MGFs (for $M_c = 1 \rm ~M_{\odot}$).
For the X-ray band, \textit{Einstein Probe}, \textit{Chandra,} and \textit{XMM-Newton} cannot detect those radiations with $d_L \gtrsim 100 {\rm~kpc}$. However, if the distance is very close, of $\sim 1 ~\rm kpc$ for example, the persistent preburst radiation produced by a bow shock might become dominant and could potentially be detected by X-ray telescopes.

\subsection{Radiation properties for different parameters}
\label{sect:param}

In this work, we consider that a magnetar and a main sequence star are in a binary system, and analyze the radiation from the interaction between magnetar bursts and the companion star. Such binary systems are relatively common in the Universe for the following reasons: (1)
The fraction of magnetars in young NSs could be about 40\% based on Galactic observations~\citep{Beniamini_2019MNRAS.487.1426B}, although the magnetic fields of magnetars would significantly decay after $\sim(10^3-10^4)~{\rm yr}$.
(2) According to the catalog of Galactic X-ray binaries~\citep{Avakyan_2023A&A...675A.199A,Fortin_2023A&A...671A.149F} (see also in Table.~1 of \cite{Xia_2023ApJ...957....1X}), for a binary star system that contains a NS (likely as the central engine of FRBs) and a high-mass companion with $M_c \sim 10 - 100 ~ M_\odot$ (a low-mass companion $M_c \sim 0.1 - 7 ~ M_\odot$), the range of its orbital period is $1 - 10^3~$d ($0.01-10 ~$d). Such parameter ranges are consistent with the above discussions.

\begin{figure*}
        \centering
        \subfigure[]{
                \includegraphics[scale=0.6]{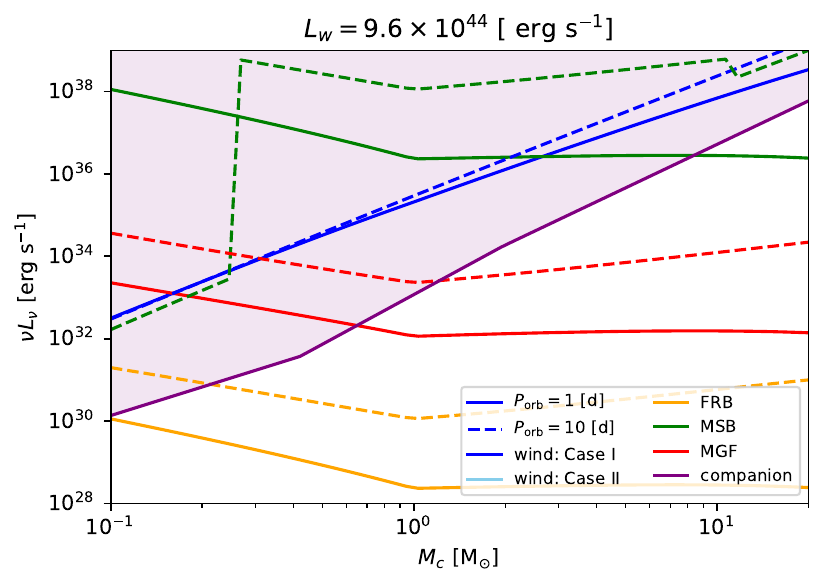}
                \label{fig:direct_radiation_a}
        }%
        \subfigure[]{
                \includegraphics[scale=0.6]{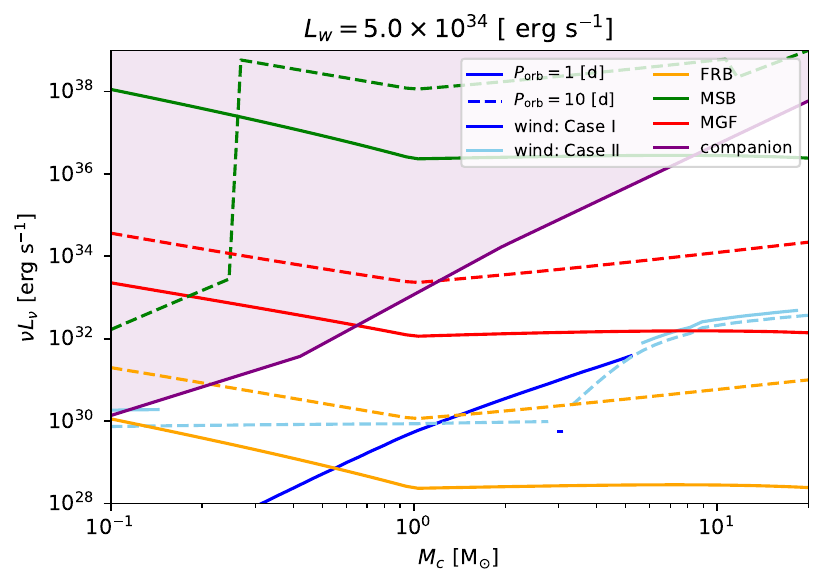}
                \label{fig:direct_radiation_b}
        }%
        
        \subfigure[]{
                \includegraphics[scale=0.6]{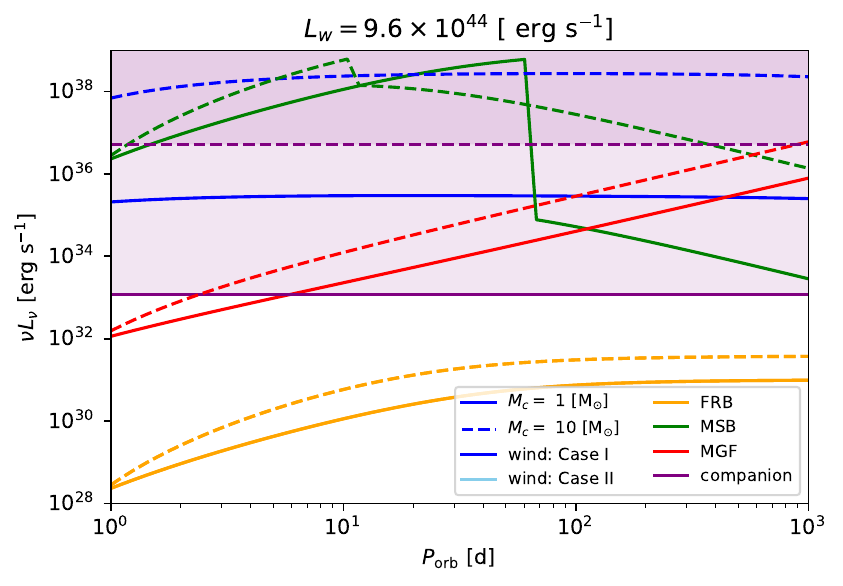}
                \label{fig:direct_radiation_c}
        }%
        \subfigure[]{
                \includegraphics[scale=0.6]{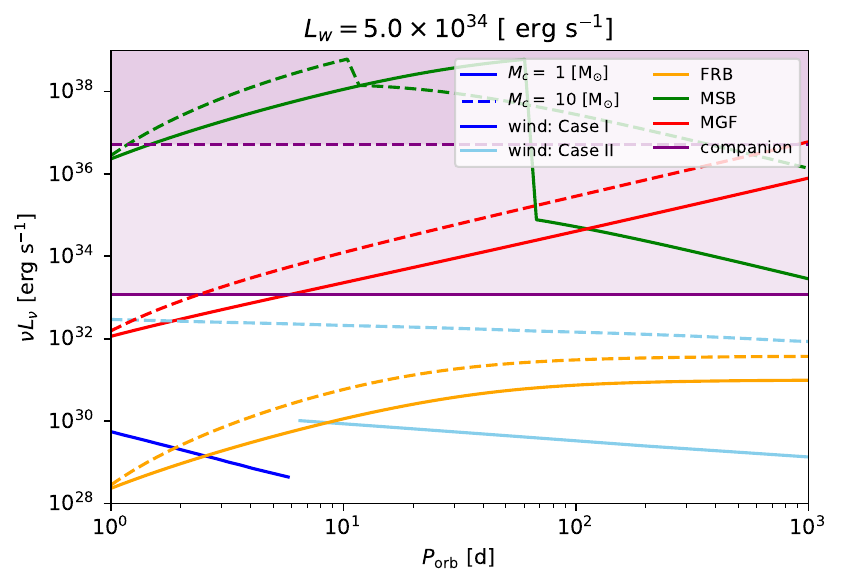}
                \label{fig:direct_radiation_d}
        }%
        \caption{Comparison of the re-emission from bursts and the persistent preburst radiation with $L_w = 9.6 \times 10^{44} {\rm~erg~s^{-1}}$ (left panels) and $L_w = 5.0 \times 10^{34} {\rm~erg~s^{-1}}$ (right panels) for different $P_{\rm orb}$ and different $M_c$. Different colors represent different mechanisms, as indicated in the caption. We note that we show both Case I (blue lines) and Case II (light blue lines) for the persistent preburst radiation. The colored region means $\nu L_\nu > L_c$, where the radiation will not be outshined by $L_c$. In the top panel, we show the variation of the peak luminosity $\nu L_\nu$ with $M_c$ for a given $P_{\rm orb} = 1$ and $10{\rm~d}$. In the bottom panel, we show the variation of $\nu L_\nu$ with $P_{\rm orb}$ for a given $M_c = 1$ and $10 {\rm~M_{\odot}}$.}
        \label{fig:direct_radiation}
\end{figure*}

\begin{figure*}
        \centering
        \subfigure[]{
                \includegraphics[scale=0.6]{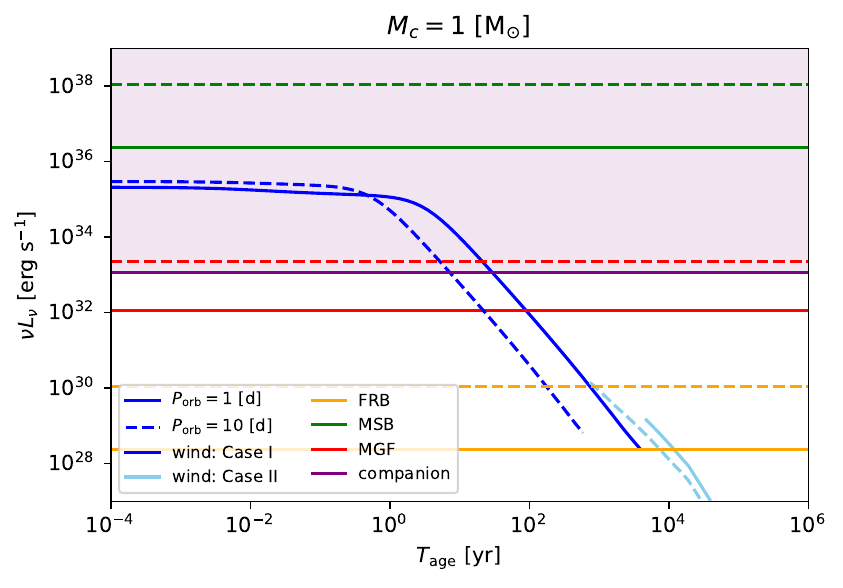}
                \label{fig:vLw1}
        }%
        \subfigure[]{
                \includegraphics[scale=0.6]{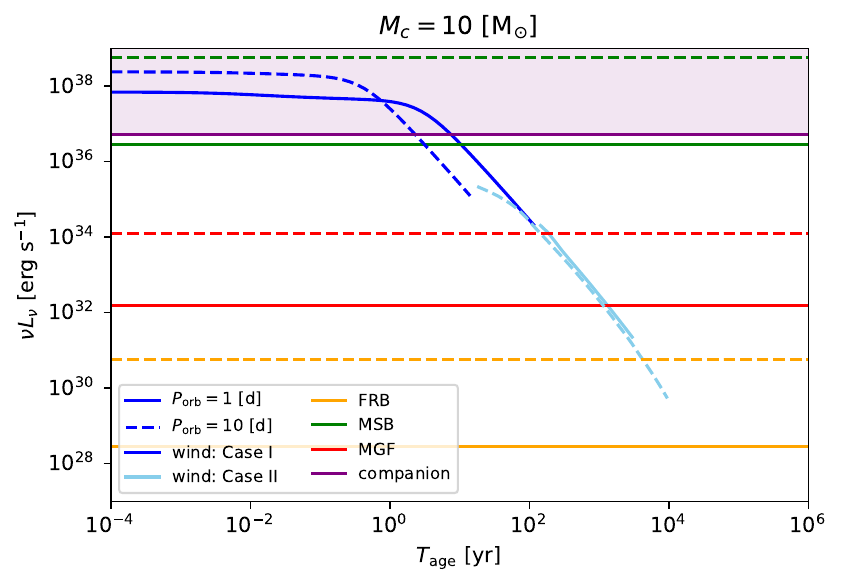}
                \label{fig:vLw2}
        }%
        \caption{Persistent preburst radiation varied with the age of the magnetar corresponding to the luminosity of the magnetar wind $L_w$ for $M_c = 1 {\rm~M_{\odot}}$ (the left panel) and $10 {\rm~M_{\odot}}$ (the right panel). Similar to Fig.~\ref{fig:direct_radiation}, the colored region means $\nu L_\nu > L_c$, and differently colored lines show different mechanisms. The solid lines show the peak luminosity for $P_{\rm orb} = 1 {\rm~d}$, while the dashed lines show that for $P_{\rm orb} = 10 {\rm~d}$.}
        \label{fig:variedLw}
\end{figure*}

We discuss the parameter dependence of the radiation from different mechanisms including the persistent preburst radiation (blue lines and light blue lines), the re-emission from FRBs (yellow lines), MSBs (green lines), and MGFs (red lines), and the radiation from the companion itself (purple lines), as shown in Fig.~\ref{fig:direct_radiation} and Fig.~\ref{fig:variedLw}. 
We show their peak luminosity in these figures, and the colored region shows the parameter space where the luminosity is larger than $L_c$. The blue lines show the persistent preburst radiation from the magnetar wind directly interacting with the companion (Case I), and the light blue lines represent that produced by the bow shock (Case II). The break between the blue line and the light blue line is due to the transition from Case I to Case II. We note that here $d > R_c$ and $d - R_c > d_{\rm  Roche}$ is satisfied for the parameter ranges we choose, where $d_{\rm  Roche} \simeq R_c (2 M_m / M_c)^{1/3}$ is the distance of the Roche limit. 

\subsubsection{Radiation properties for different $M_c$ and $P_{\rm orb}$}

In Fig.~\ref{fig:direct_radiation}, we set $L_w = 9.6 \times 10^{44} {\rm~erg~s^{-1}}$ for the typical parameters, corresponding to a newborn magnetar, and set $L_w = 5.0 \times 10^{34} {\rm~erg~s^{-1}}$ , which corresponds to a relatively old magnetar with $T_{\rm age} = 10^3 {\rm~yr}$ for example. 
We scan the parameter space of $M_c = 0.1 - 20 {~\rm M_{\odot}}$ and $P_{\rm orb} = 1 - 1000 {\rm~d}$. 

In Fig.~\ref{fig:direct_radiation_a} and Fig.~\ref{fig:direct_radiation_b}, we show the peak luminosity of the radiation for $M_c = 0.1 - 20 {\rm~M_{\odot}}$ with $P_{\rm orb} = 1, \ 10 {\rm~d}$. We note that the discontinuity of the solid green line represents the transition point after which the extra limitation $r'_y - r'_x \leq c \Delta t_{\rm burst}$ is not satisfied and we use $r'_y = r'_x + c \Delta t_{\rm burst}$.
We can see that for the companion with a small $M_c \lesssim 0.6 {\rm~ M_{\odot}}$, the re-emission from MGFs and MSBs can be larger than $L_c$. Only with $M_c \lesssim 0.21 {\rm~ M_{\odot}}$ and $P_{\rm orb} = 10~$d can the re-emission from FRBs be larger than $L_c$.
In Fig.~\ref{fig:direct_radiation_a}, we set $L_w = 9.6 \times 10^{44} {\rm~erg~s^{-1}}$ corresponding to Case I. 
For $P_{\rm orb} = 1 {\rm~d}$ ($10 {\rm~d}$), the re-emission from MSBs can be greater than the persistent preburst radiation with $0.25 {\rm~ M_{\odot}} \lesssim M_c \lesssim 9.8 {\rm~ M_{\odot}}$ ($M_c \lesssim 2.5 {\rm~ M_{\odot}}$), and that from MGFs can be greater than the persistent preburst radiation with $M_c \lesssim 0.16 {\rm~ M_{\odot}}$ ($0.29 {\rm~ M_{\odot}}$).
In Fig.~\ref{fig:direct_radiation_b}, we use $L_w = 5.0 \times 10^{34} {\rm~erg~s^{-1}}$, and the persistent preburst radiation would be outshined by the radiation from the companion itself. 

In Fig.~\ref{fig:direct_radiation_c} and Fig.~\ref{fig:direct_radiation_d}, we show the peak luminosity for $P_{\rm orb} = 1 - 1000 {\rm~d}$ with $M_c = 1, \ 10 {\rm~M_{\odot}}$. We can see that $\nu L_\nu$ converges as $P_{\rm orb}$ increases, and the re-emission from MGFs and MSBs is greater than that from FRBs for all the selected values of $P_{\rm orb}$. The re-emission from MGFs is larger than that from MSBs only with a relative large $P_{\rm orb}$ for $M_c = 1 - 10 ~{\rm M_{\odot}}$. For the re-emission from FRBs (MGFs), it will be outshined by $L_c$ with $M_c = 1-10 {\rm~M_{\odot}}$ ($10 {\rm~M_{\odot}}$). 
We note that our results do not imply that the re-emission from FRBs must be outshined, because if the transformation efficiency is higher, it has the possibility to be higher than the radiation of the companion star itself.
In Fig.~\ref{fig:direct_radiation_c}, similar to Fig.~\ref{fig:direct_radiation_a}, we set $L_w = 9.6 \times 10^{44} {\rm~erg~s^{-1}}$.
For the companion with $M_c = 1 {\rm~M_{\odot}}$, the re-emission from MGFs (MSBs) is dominant for $P_{\rm orb} \gtrsim 440 {\rm~d}$ ($\lesssim 60 {\rm~d}$).
For the companion with $M_c = 10 {\rm~M_{\odot}}$, the persistent preburst radiation would be dominant for all the values of $P_{\rm orb}$ we choose here, except for $5.2 {\rm~d} \lesssim P_{\rm orb} \lesssim 9.2 {\rm~d,}$ where the re-emission from MSBs is dominant.
In Fig.~\ref{fig:direct_radiation_d}, similar to Fig.~\ref{fig:direct_radiation_b}, the re-emission from MSBs and MGFs would be greater than the persistent preburst radiation with all the $P_{\rm orb}$ we choose. 

The light blue lines in Fig.~\ref{fig:direct_radiation_d} show the persistent preburst radiation from the bow shock. We find that the radiation of the bow shock will change from the fast-cooling case to the slow-cooling case with increasing distance $d,$ corresponding to increasing orbital period $P_{\rm orb}$, and the peak flux in the slow-cooling case is inversely proportional to the distance $d$. As $d$ continues to increase, there will be $\gamma_c > \gamma_M$, where the cooling will no longer be important, and so the radiation from the bow shock is weak.

\subsubsection{Radiation properties for different $L_w$}

According to Sect.~\ref{sect:model}, we know that the persistent preburst radiation varies with the luminosity of the magnetar wind $L_w$. Therefore, in Fig.~\ref{fig:variedLw}, we show the persistent preburst radiation as a function of $L_w$, and compare it with the re-emission from bursts and the companion itself.  
For $M_c = 1 {\rm~M_{\odot}}$, we find the persistent preburst radiation would be larger than $L_c$ with $T_{\rm age} \lesssim 36~$yr ($8.9~$yr) for $P_{\rm orb} = 1 {\rm~d}$ ($10 {\rm~d}$).
The persistent preburst radiation would be smaller than that of the re-emission from MSBs with all the values of $T_{\rm age}$ we choose for $P_{\rm orb} = 1-10 {\rm~d}$. The re-emission from MGFs can be larger than the persistent preburst radiation with $T_{\rm age} \gtrsim 5.6 {\rm~yr}$ for $P_{\rm orb} = 10 {\rm~d}$, and the other re-emission would be outshined by $L_c$.
For $M_c = 10 {\rm~M_{\odot}}$, the persistent preburst radiation will be larger than $L_c$ with $T_{\rm age} \lesssim 8.9~$yr ($2.8~$yr) for $P_{\rm orb} = 1 {\rm~d}$ ($10 {\rm~d}$).
The re-emission from MSBs will be larger than the persistent preburst radiation with all the values of $T_{\rm age}$ we choose for $P_{\rm orb} = 10 {\rm~d}$, and the other re-emission is outshined by $L_c$.

\section{Summary and discussion}

In this work, we studied the radiation processes from the interaction between magnetar bursts and the main sequence star (the companion) in a binary system. We considered the possible radiation generated by the magnetar wind interacting with the companion, or the companion wind as the background of the re-emission from bursts (including FRBs, MSBs, and MGFs) interacting with the companion. For the persistent preburst radiation, we considered two possible scenarios based on $L_w$. 
For a magnetar wind with a relatively small $L_w$, the magnetar wind will interact with the companion wind, generating the bow shock; whereas for a magnetar wind with a large $L_w$, it will directly interact with the companion itself. 

We calculated the spectra and light curves for the companion with $M_c = 1 {\rm~M_{\odot}}$ and $10 {\rm~M_{\odot}}$ in a binary system with $P_{\rm orb} = 10 {\rm~d}$. 
We find that only when the luminosity of the magnetar wind is relatively large ---for example, $9.6 \times 10^{44} {~\rm erg ~s^{-1}}$, corresponding to a newborn magnetar with $P_0 = 10^{-2} {\rm~s}$ and $B_{\perp} = 10^{15} {\rm~G}$--- can the persistent preburst radiation be outshined by $L_c$, and this mainly happens at optical and UV bands. 
The re-emission from MGFs with $M_c = 1 {\rm~M_{\odot}}$ is slightly larger than that of $L_c$, the re-emission from MSBs with $M_c = 1-10 {\rm~M_{\odot}}$ is $\sim 1-4$ orders of magnitude larger than that of $L_c$, and that from FRBs with $M_c = 1-10 {\rm~M_{\odot}}$ is outshined by $L_c$. We note that the re-emission from FRBs with a larger transformation efficiency $\eta$ could still potentially be observed. The re-emission from the bursts is mainly at the optical, UV, and X-ray bands, with a duration of $\sim 0.1 - 10^5 {\rm~s}$. 
For the optical band, we find that the Vera C. Rubin Observatory can detect the persistent preburst radiation with $M_c = 1-10 {\rm~M_{\odot}}$ and the re-emission from MGFs with $M_c = 1 {\rm~M_{\odot}}$. 
For the X-ray band, \textit{Einstein Probe}, \textit{Chandra,} and \textit{XMM-Newton} cannot detect the radiations we study with $d_L \gtrsim 100~$kpc.
However, in the high-energy band $\gtrsim 10^{17} {\rm~Hz}$, the persistent preburst radiation produced by a bow shock can become dominant, and could potentially be detected when the distance is very close $\sim 1 {\rm~kpc}$.

We also scanned the parameter space of $M_c$, $P_{\rm orb}$, and $L_w$ to find which mechanism is dominant, and we find that the persistent preburst radiation and the re-emission from bursts can be larger than the radiation from the companion itself with a suitable parameter. Also, the re-emission from FRBs, MSBs, and MGFs can be greater than the persistent preburst radiation for a relatively old magnetar.
For FRBs, when $M_c \lesssim 0.2 {\rm ~M_{\odot}}$, its re-emission can be greater than $L_c$ and the persistent preburst radiation of $L_w \approx 5.0 \times 10^{34} {~\rm erg ~s^{-1}}$ with $P_{\rm orb} = 10 {\rm~d}$.
For MSBs and MGFs, their re-emission can be larger than $L_c$ and the persistent preburst radiation of $L_w \approx 5.0 \times 10^{34} {~\rm erg ~s^{-1}}$ with $M_c \lesssim 0.6 {\rm ~M_{\odot}}$. 
For the persistent preburst radiation with $L_w \approx 9.6 \times 10^{44} {~\rm erg ~s^{-1}}$, which corresponds to a newborn magnetar with $P_0 = 10^{-2} {\rm~s}$ and $B_{\perp} = 10^{15} {\rm~G}$, it would be dominant with a relatively large $M_c$ ($\gtrsim 1-10 ~M_{\odot}$) and a moderate $P_{\rm orb}$ ($\sim 10-100~$d).
We note that here we compare the peak luminosity for each radiation mechanism, and in addition to the persistent preburst radiation produced by the bow shock (Case II), the other radiation mechanism is the black-body radiation. 
Therefore, in high-energy bands, such as X-ray and $\gamma$-ray bands, the persistent preburst radiation produced by a bow shock might become dominant.

In the above calculations, we focus on the two main radiations, including the persistent preburst radiation from the magnetar wind and the re-emission from bursts. 
However, there are some additional mechanisms, including the IC scattering process between the electrons in the magnetar wind and the photons from the re-emission, and the pair annihilation from the electrons in the companion and the positrons in the magnetar wind (some details of these processes are discussed in the Appendix).
We find the peak luminosity from the re-emission can be scattered to a larger frequency with the IC scattering process.
We also find that the pair annihilation would exist, but the photons generated by this process would be scattered by the particles in the companion star due to the optically thick surface of the  companion, which cannot be detected.

\begin{acknowledgements}
We thank the anonymous referee for helpful comments to improve this paper. We thank the helpful discussions with Bing Zhang, Jia Ren, and Fa-Yin Wang, and we thank the useful information given by Xin-Lian Luo, Yue Wu, Bo-Yang Liu, Zhen-Yin Zhao, Qian-Qian Zhang, and Lu-yao Jiang.
D.M.W is supported by the National Natural Science Foundation of China (No. 11933010, 11921003, 12073080, 12233011) and the Strategic Priority Research Program of the Chinese Academy of Sciences (grant No. XDB0550400). Z.G.D is supported by the National SKA Program of China (grant No. 2020SKA0120300) and National Natural Science Foundation of China (grant No. 12393812). Y.P.Y is supported by the National Natural Science Foundation of China grant No.12003028 and the National SKA Program of China (2022SKA0130100). 
\end{acknowledgements}

%
\bibliographystyle{aa} 
\bibliography{yjwei} 
%

\begin{appendix}
\section{Pair annihilation between electrons in companion and positrons in relativistic wind} 

\begin{figure}
        \centering
        \subfigure[]{
                \includegraphics[scale=0.6]{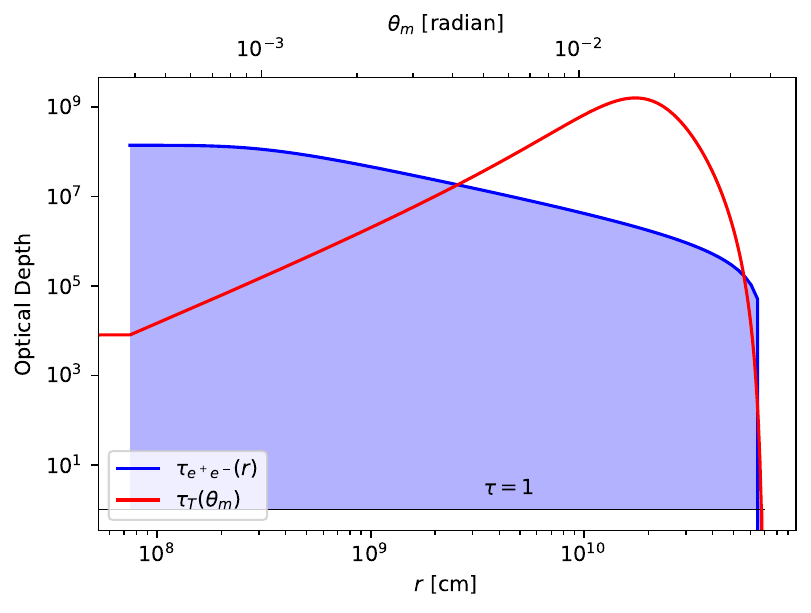}
        }%
        
        \caption{Optical depth for a solar-like companion with $M_c = 1 {\rm~M_{\odot}}$. The blue line shows the optical depth for the pair annihilation at different radii, and the colored region means $\tau_{e^+e^-} > 1$. The red line represents the optical depth of absorption at various viewing angles $\theta_m \in (0, \theta_{m2})$. The corresponding parameters we used are $\gamma_w = 10^4$ and $P_{\rm orb} = 10 {\rm ~d}$.}
        \label{fig:pair}
\end{figure}

The pair annihilation is one of the possible heating mechanisms due to the very high density of the electrons in the companion, in which an electron (from the companion) and a positron (from the magnetar wind) annihilate, resulting in the creation of two photons.
The cross section for this process is \citep{Jauch_1976tper.book.....J}
\begin{equation}
        \begin{aligned}
                & \sigma_{e^+e^-} = \frac{\pi r_0^2}{\gamma_w + 1} \left[ \frac{\gamma_w^2 + 4 \gamma_w + 1}{\gamma_w^2 - 1} {\rm ln} \left( \gamma_w + \sqrt{\gamma_w^2 - 1}\right) - \frac{\gamma_w + 3}{\sqrt{\gamma_w^2 - 1}}  \right].
        \end{aligned}
\end{equation}
The corresponding optical depth $\tau_{e^+e^-} \sim n_c \sigma_{e^+e^-} r_b$ can be calculated, where $n_c \simeq \rho_c / m_p$ is the number density of electrons in the companion. The optical depth can be much larger than $1$ due to the large $n_c$, as shown in the blue line in Fig.~\ref{fig:pair}, which means that the possibility of pair annihilation would be large and the radiation from this process cannot be neglected. However, the emitting photons might be scattered by the particles in the companion star due to the optically thick companion surface. We then calculate the corresponding optical depth $\tau_{T} \sim \kappa \rho_c l (\theta),$ considering the Thomson scattering, where $l (\theta)$ is the length of the companion star at a given $\theta_m$ ($\theta_m$ is the angle between the line connecting the two stars and the line connecting the magnetar and the point we studied). The descriptions of $\theta_m$ and $\theta_{m2}$ are presented in Fig.~\ref{fig:windmodel}. According to the red line in Fig.~\ref{fig:pair}, we can see that the radiation could only be potentially observed with a $\theta_m$ of around $\theta_{m2} \approx 0.038 {\rm~rad}$ (where $\tau_T < 1$). Therefore, the radiation from the pair annihilation can be neglected, and it contributes to the heating of the companion star by the magnetar wind.

\section{Inverse Compton between electrons in relativistic wind and photons from re-emission}
\label{sect:IC}

\begin{figure}
        \centering
        \subfigure[]{
                \includegraphics[scale=0.6]{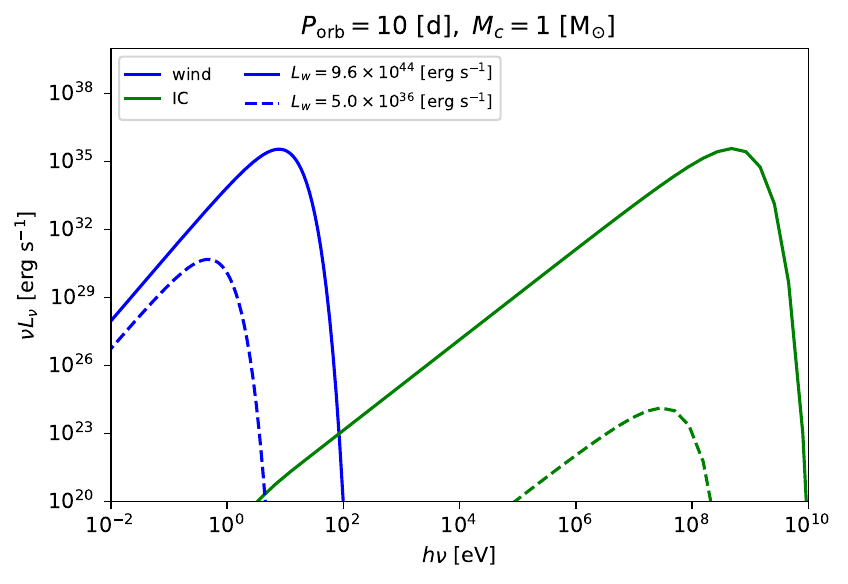}
        }%
        
        \caption{Spectra of the persistent preburst radiation for Case I involving the IC process with $T_{\rm age} = 10^2 {\rm~yr}$ corresponding to $L_w \simeq 5.0 \times 10^{36} {\rm~erg~s^{-1}}$ (the dashed line) and $T_{\rm age} = 0 {\rm~yr}$ corresponding to $L_w \simeq 9.6 \times 10^{44} {\rm~erg~s^{-1}}$ (the solid line). We set $P_{\rm orb} = 10 {\rm~day,}$ corresponding to $a \approx 3.9 \times 10^{11} {~\rm cm}$  and $M_c = 1 {\rm~M_{\odot}}$. The blue lines show the preburst radiation spectrum of Case I. The green line indicates the IC spectrum. The corresponding parameters we used are $p = 2.2$, $\gamma_m = 10^4$, and $\gamma_M = 10^6$.}
        \label{fig:ic}
\end{figure}

The inverse Compton (IC) scattering process may also  take place between the electrons in the relativistic wind and the photons from the re-emission. The IC spectrum for isotropic electrons and isotropic photons at $a_1 = h \nu_1 / (m_e c^2)$ (where $\nu_1$ is the emitting photon frequency) should be \citep{Blumenthal_1970RvMP...42..237B}
\begin{equation}
        \begin{aligned}
                & L_{\rm IC} \sim \nu_1 L_{\nu_1} = V_{\rm IC} h \nu_1^2 \int \int n_e (\gamma) n_{\rm ph} (\nu) \frac{d N_{\gamma, \nu}}{dt d\nu_1} d\nu d\gamma, \\
                & \frac{d N_{\gamma, \nu}}{d t d \nu_1} = \frac{2 \pi r_0^2 c}{\nu \gamma^2} \left[2 q {\rm ln} q + (1+2q)(1-q) + \frac{1-q}{2} \frac{(4a_1 \gamma q)^2}{1 + 4a_1 \gamma q} \right],
        \end{aligned}
\end{equation}
where $q = \frac{(a_1 / \gamma)^2}{4 a a_1 (1 - a_1/\gamma)}$ is a dimensionless parameter, $r_0 = e^2 / (m_e c^2)$ is the classical electron radius, and $n_e (\gamma) = n_{e,0} (1-p) \frac{\gamma^{-p}}{\gamma_M^{-p+1} - \gamma_m^{-p+1}}, \ \gamma_m < \gamma < \gamma_M$ is the differential number density of the power-law distribution electrons contributed by the magnetar wind. 
We take the photons produced by the persistent preburst radiation for Case I as the seed photon for example. We use $P_w = P_{\rm re}$ to estimate the width of the interaction region (where $P_{\rm re} \simeq L_{\rm re} / (12 \pi {\delta r}^2 c)$ is the radiation pressure), which is $\delta r \sim (a - R_c) / (1 + \sqrt{L_w / L_{\rm re}})$, as shown in Fig.~\ref{fig:windmodel1}. We then approximate that the volume of the interaction region is $V_{\rm IC} \sim \pi \delta r R_c^2$, and the number density of the photons is $n_{\rm ph} (\nu) = L_{\nu} / (4 \pi h \nu (R_c + \delta r)^2 c)$. 
We note that $n_{e,0} \simeq \dot{N} / (4 \pi (a - R_c)^2 c)$ is the number density for power-law distribution electrons, where $\dot{N} \simeq \frac{L_w}{m_e c^2} \frac{2-p}{1-p} \frac{\gamma_M^{-p+1} - \gamma_m^{-p+1}}{\gamma_M^{-p+2} - \gamma_m^{-p+2}}$ is the number of electrons per second at $a - R_c$. 

In Fig.~\ref{fig:ic}, we consider the case where the seed photon for the IC process is mainly from the persistent preburst radiation generated by Case I. Here we take the solar-like companion with $M_c = 1 {\rm~M_{\odot}}$ and the newborn magnetar corresponding to $L_w \approx 9.6 \times 10^{44} {\rm~erg ~s^{-1}}$ (the magnetar with $T_{\rm age} = 10^2 {~\rm yr}$ corresponding to $L_w \approx 5.0 \times 10^{36} {\rm~erg ~s^{-1}}$) for example. We find that the peak luminosity for the persistent preburst radiation can be scattered to a large frequency. 
For $T_{\rm age} = 10^2 {~\rm yr}$, the peak luminosity for the IC spectrum is $\approx 1.3 \times 10^{24} {~\rm erg ~ s^{-1}}$ at $\nu \approx 6.9 \times 10^{21} {~\rm Hz}$. For a newborn magnetar, the peak luminosity for IC is $\approx 3.8 \times 10^{35} {~\rm erg ~ s^{-1}}$ at $\nu \approx 1.2 \times 10^{23} {~\rm Hz}$, which is similar to that for the persistent preburst radiation due to the large number density of electrons.
Considering the possible two-photon pair production process between the photons from the IC scattering process ($\nu_{\rm IC}$) and the lower energy photons ($\nu_{\rm seed}$), the frequency of the lower energy photons should satisfy $\nu_{\rm seed} \geq (m_e c^2 / h)^2 \nu_{\rm IC}^{-1} \simeq 1.3 \times 10^{17} ~{\rm Hz } ~\nu_{\rm IC, 24}^{-1} $, which is much larger than the peak frequency of the radiation from the solar-like companion itself $\approx 3.4 \times 10^{14} {\rm~Hz}$ with $T_{c, \rm eff} \approx 5770 {\rm~K}$. In this way, here we neglect this effect. For Case II of the persistent preburst radiation, the IC scattering process can also scatter the seed photons to a relatively high frequency of $\sim 10^{24} \rm~Hz$.

In addition, we investigate the impact of the repeated scatterings involving relativistic electrons in purely scattering scenarios, by calculating the Compton $Y$ parameter~\citep{Rybicki_1985rpa..book.....R}
\begin{equation}
    Y \simeq \frac{64}{3} \left< \gamma^2 \right> \max (\tau_{\rm es}, \tau^2_{\rm es}),
\end{equation}
where $\tau_{\rm es} = \rho_e \kappa \delta r$ is the scattering optical depth, and $\rho_e = m_e n_e$ is the density of the electrons. We find that the value of $Y$ is about $4.8 \times 10^{-5}$ ($4.1 \times 10^{-3}$) for the magnetar with $T_{\rm age} = 10^3~$yr (the newborn magnetar), which is smaller than $1,$ indicating that the repeated scattering process is not important. 

\end{appendix}

\end{CJK*}
\end{document}